\documentclass[11pt,letterpaper,twoside,english]{article}
\usepackage[latin9]{inputenc}
\usepackage{amsmath}
\usepackage{amssymb}
\usepackage{graphicx}
\usepackage{esint}

\makeatletter

\newcommand{\lyxmathsym}[1]{\ifmmode\begingroup\def\b@ld{bold}
  \text{\ifx\math@version\b@ld\bfseries\fi#1}\endgroup\else#1\fi}

\@ifundefined{date}{}{\date{}}
\usepackage{times}\usepackage{fancyhdr}

\def\cleardoublepage{\clearpage\if@twoside \ifodd\c@page\else% 
    \hbox{}% 
    \thispagestyle{empty}%
    \newpage% 
    \if@twocolumn\hbox{}\newpage\fi\fi\fi}

\def\figurename{Figure}

\renewcommand{\fnum@figure}[1]{\figurename~\thefigure.}

\def\tablename{Table}

\renewcommand{\fnum@table}[1]{\tablename~\thetable.}

\usepackage{babel}

\makeatother

\usepackage{babel}
\begin{document}

\title{\vskip 0.45in \textbf{\textsc{Configurations of Structural Defects
in Graphene and Their Effects on Its Transport Properties }}}

\author{\textbf{\textit{T. M. Radchenko$^{1,}$}}%
\thanks{\textbf{\textit{E-mail address: tarad@imp.kiev.ua}}%
}\textbf{\textit{~, V. A. Tatarenko$^{1,}$}}%
\thanks{\textbf{\textit{E-mail address: tatar@imp.kiev.ua}}%
}\textbf{\textit{~, I. Yu. Sagalianov$^{2,}$}}%
\thanks{\textbf{\textit{E-mail address: isagal@ukr.net}}%
}\textbf{\textit{ }}\\
 \textbf{\textit{ and Yu. I. Prylutskyy$^{2,}$}}%
\thanks{\textbf{\textit{E-mail address: prylut@ukr.net}}%
}\\
 \textbf{\textit{ $^{1}$G. V. Kurdyumov Institute for Metal Physics,
N.A.S.U., }}\\
 \textbf{\textit{Kyiv, Ukraine}}\\
 \textbf{\textit{ $^{2}$Taras Shevchenko National University of Kyiv,
}}\\
 \textbf{\textit{Kyiv, Ukraine}}}

\maketitle
\thispagestyle{empty} \setcounter{page}{1}

%\vspace{0in}

\begin{abstract}
\noindent The chapter combines analytical (statistical-thermodynamic
and kinetic) with numerical (Kubo--Greenwood-formalism-based) approaches
used to ascertain an influence of the configurations of point (impurities,
vacancies) and line (grain boundaries, atomic steps) defects on the
charge transport in graphene. Possible substitutional and interstitial
graphene-based superstructures are predicted and described. The arrangements
of dopants over sites or interstices related with interatomic-interaction
energies governing the configurations of impurities. Depending on
whether the interatomic interactions are short- or long-range, the
low-temperature stability diagrams in terms of interaction-energy
parameters are obtained. The dominance of intersublattice interactions
in competition with intrasublattice ones results in a nonmonotony
of ordering-process kinetics. Spatial correlations of impurities do
not affect the electronic conductivity of graphene for the most important
experimentally-relevant cases of point defects, neutral adatoms and
screened charged impurities, while atomic ordering can give rise in
the conductivity up to tens times for weak and strong short-range
potentials. There is no ordering effect manifestation for long-range
potentials. The anisotropy of the conductivity along and across the
line defects is revealed and gives rise in the conductivity of graphene
with correlated line defects as compared with the case of random ones.
Simultaneously correlated (and/or ordered) point and line defects
in graphene can give rise in the conductivity up to hundreds times
vs. their random distribution. On an example of different B or N doping
configurations in graphene, results from the Kubo--Greenwood approach
are compared with those obtained from DFT method. 
\end{abstract}
% ------------ [Running Heads - pagina 2] -------------------------------------------------
\pagestyle{fancy} \fancyhead{} \fancyhead[EC]{T. M. Radchenko, V. A. Tatarenko, I. Yu. Sagalianov et al.}
\fancyhead[EL,OR]{\thepage} \fancyhead[OC]{Configurations of Structural Defects in Graphene and
Their Effects ...} \fancyfoot{} \global\long\def\headrulewidth{0.5pt}
 %-------------------------------------------------------------------------------


\vspace{0.08in}

\noindent %\noindent \vspace{0.08in}

\section*{Introduction}

Pure and structurally perfect graphene has shown outstanding electronic
phenomena such as ballistic electron propagation with extremely high
carrier mobilities \cite{Novoselov2004} or the quantum Hall effect
at room temperature \cite{Novoselov2007}. However, the absence a
band-gap in pristine graphene makes its current--voltage characteristic
symmetrical with respect to the zero-voltage point and thereby does
not allow switching of graphene-based transistors with a high on--off
ratio. There are different ways to induce a band-gap in graphene,
particularly by the introduction of impurities (point defects) \cite{Krasheninnikov2011}.

Generally, different types of defects are always present in crystals
due to the imperfection of material production processes. Such lattice
imperfections strongly affect different properties in solids. Defect
in bulk crystals are investigated for many years, while graphene-based
materials are considered only recently. Investigation of its transport
properties and understanding factors that affect its conductivity
represent one of the central directions of graphene research \cite{CastroNetoreview,PeresReview,DasSarmaReview}.
This is motivated by both fundamental interest to graphene's transport
properties as well as by potential applications of this novel material
for electronics. It is commonly recognized that the major factors
determining the electron mobility in graphene are long-range charged
impurities trapped on the substrate and strong resonant short-range
scatterers due to adatoms covalently bound to graphene \cite{PeresReview}.
The nature of impurity atoms, acting as the scatterers, is directly
reflected in the dependence of the conductivity on the electron density,
$\sigma=\sigma(n_{e})$, and therefore investigation of this function
constitutes the focus of experimental and theoretical research \cite{PeresReview,DasSarmaReview}.
Dopant atoms change the band structure strongly dependent on atomic
order and, consequently, provide a tool to govern and even to control
electrical conductivity of the graphene-based materials.

Defects, playing role of disorder, can be not always random and stationary,
migrating with a certain mobility governed by the activation barrier
and temperature \cite{Krasheninnikov2011}. Such migration and relaxation
to the equilibrium state as well as the features of the growth technology
can result in a correlation or even an ordering in the configuration
of point or/and line defects. Experimental observation of correlation
in the spatial distribution of disorder have been reported in Ref.
\cite{YanFuhrer}, where authors addressed enhancement of the conductivity
to the effect of correlation between the potassium atoms doping the
graphene. This conclusion, in turn, was based on the theoretical predictions
in Ref. \cite{Li} that the correlations in the position of long-range
scatterers strongly enhances the conductivity. It should be noted
that the approach in Ref. \cite{Li} is based on the standard Boltzmann
approach within the Born approximation. However, the applicability
of the Born approximation for graphene has been questioned in Ref.
\cite{Klos}, where it was shown that predictions based on the standard
semi-classical Boltzmann approach within the Born approximation for
the case of the long-range Gaussian potential are in quantitative
and qualitative disagreement with the exact quantum-mechanical results
in the parameter range corresponding to realistic systems. Therefore
it is of interest to study the effect of spatial correlation of dopant
atoms by exact numerical methods.

Correlation of extended defects, which act as line scatterers, has
been also experimentally revealed \cite{Kuramochi2012}. Particularly,
it concern the line defects in epitaxial and chemically-vapor-deposited
(CVD) graphene, where they manifest correlation in their orientation
and can be even almost parallel to each other due to the growth technique
\cite{Kuramochi2012,Gunther2011,Held2012,Ni2012}. Such correlation
leads to the anisotropy of diffusivity and conductivity in different
directions of graphene sheets \cite{Kuramochi2012,Ni2012,Yakes2010}.

In the present chapter, we consider different (re)distributions of
point and one-dimensional (1\textit{D}) defects in graphene lattice
and then ascertain how do their configurations affect the diffusivity
and hence conductivity of charge carriers. To do it, we combine analytical
(statistical-thermodynamic along with kinetic) approaches and numerical
(quantum-mechanical) calculations. An advantage of analytical method
is account of interatomic interactions of all (but not only commonly
assumed nearest-neighbor) atoms in the system. Numerical (Kubo--Greenwood-formalism-based)
calculations are especially suited to treat large graphene sheets
containing millions of atoms, i.e. with dimensions approaching realistic
systems (here, as well as in Refs. \cite{Radchenko et al. 1,Radchenko et al. 2},
we treat systems having the sizes of $1700\times1000$ and $3400\times2000$
sites corresponding to $210\times210$ and $420\times420$ nm$^{2}$).
In case of the point defects, we consider random, correlated, and
ordered distributions of dopant atoms, calculate the conductivity
for each case, and compare obtained results. Line defects are also
considered as randomly distributed and orientationally correlated,
i.e. with a prevailing direction in their orientation.

\section*{Substitutional and Interstitial Graphene-Based Superstructures: Statistical-Thermodynamic
Approach}

Let us consider possible ordered distributions of impurity atoms over
the sites and interstices of honeycomb lattice, namely, graphene-based
substitutional and interstitial (super)structures, which are stable
against the formation of antiphase boundaries (or splitting up onto
antiphase domains).

\subsection*{Substitutional Superlattices}

\begin{figure*}[t]
\includegraphics[width=1\textwidth]{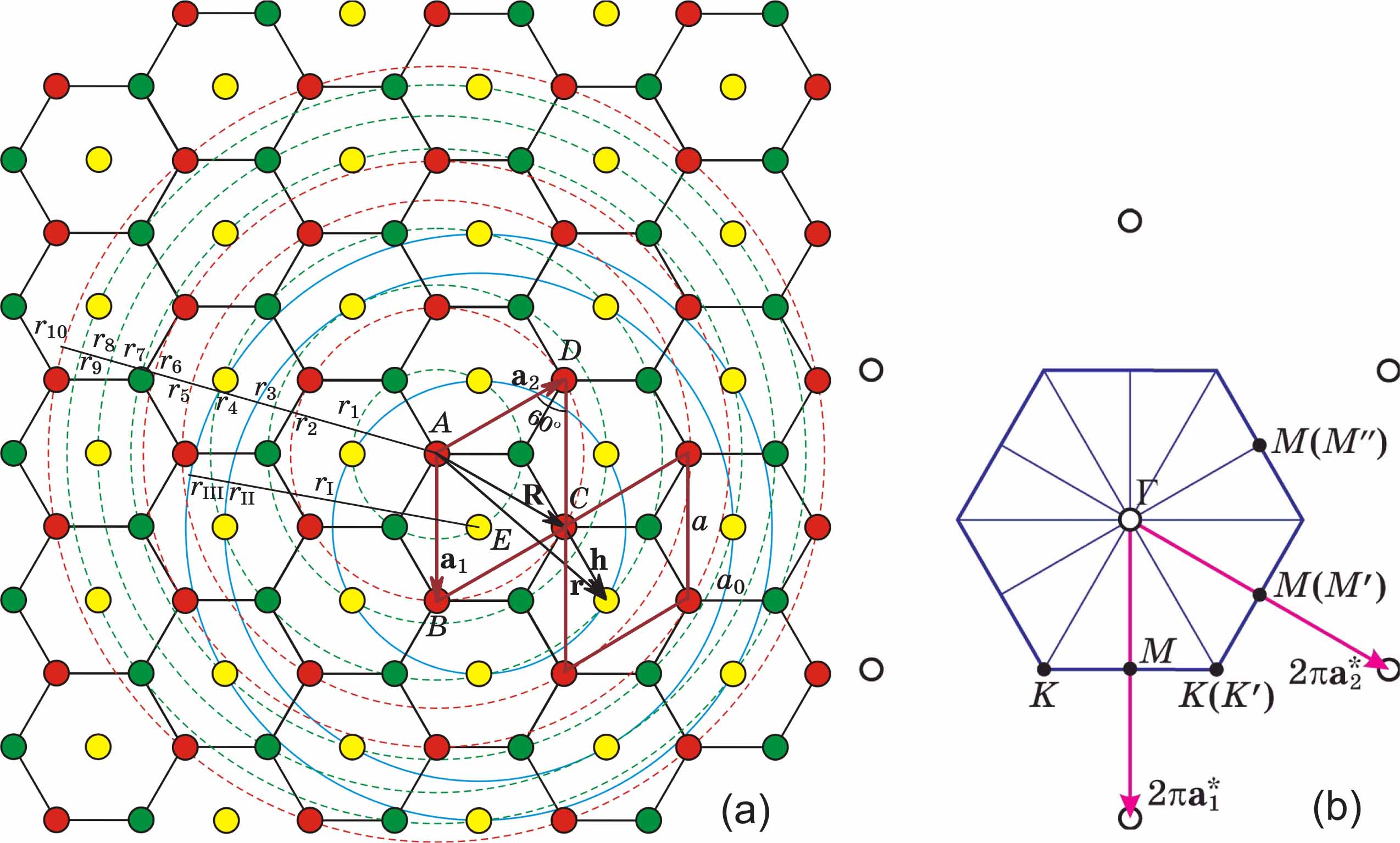}

\caption{(Color online) (a) Crystal lattice of graphene. Here, $ABCD$ is a
primitive unit cell; $\mathbf{a}_{1}$ and $\mathbf{a}_{2}$ are the
basis translation vectors of the lattice; $a$ is the lattice translation
parameter; $a_{0}$ is a distance between the nearest-neighbor sites;
circles (with radii $r_{1}$, $r_{2}$, ..., $r_{10}$, and $r_{I}$,
$r_{II}$, $r_{III}$) denote the first ten substitutional (dashed
line) and three interstitial (solid line) coordination shells (zones)
with respect to the origin (at $A$ site) of the oblique coordinate
system and interstice $E$, respectively. (b) The first Brillouin
zone ($BZ$) of the reciprocal space of graphene lattice, where $\Gamma$,
$M$, $K$ are its high-symmetry points; $\mathbf{a}_{1}^{*}$ and
$\mathbf{a}_{2}^{*}$ are the basis translation vectors of two-dimensional
reciprocal lattice.}

\label{Fig_Lattice} 
\end{figure*}

\begin{figure*}[t]
\includegraphics[width=1\textwidth]{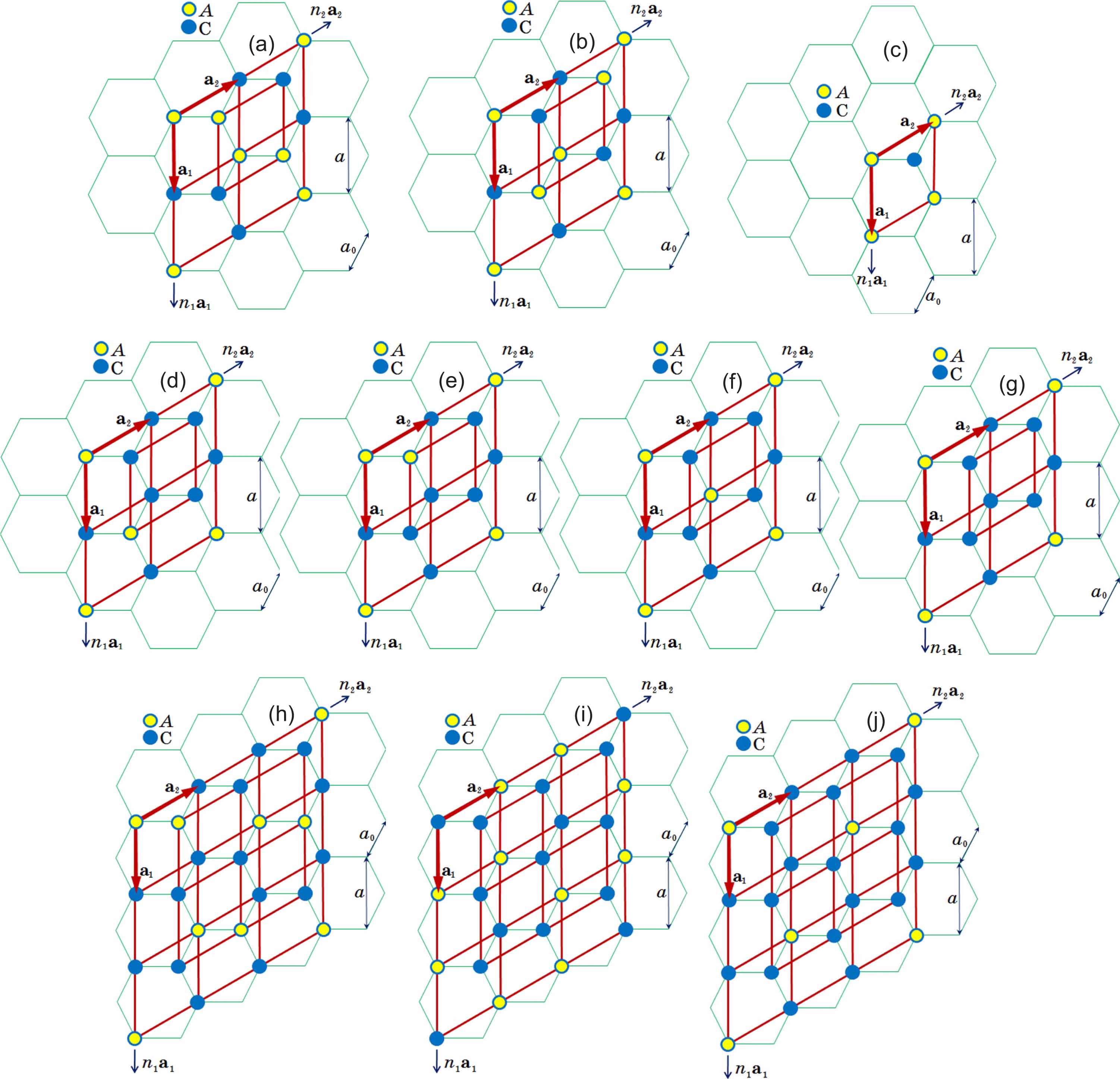}

\caption{(Color online) Primitive unit cells of substitutional superstructures
with stoichiometries $1/2$ (a)--(c), $1/4$ (d)--(f), $1/8$ (g),
$1/3$ (h)--(i), $1/6$ (j).}

\label{Fig_Superstructures_substitutional} 
\end{figure*}

Ordered distributions of substitutional atoms \textit{A} over the
sites of honeycomb lattice at the stoichiometries $c_{st}=1/2$, (\textit{$\mathrm{C}A$}),
$1/4$ (\textit{$\mathrm{C}_{3}A$}), $1/8$ (\textit{$\mathrm{C}_{7}A$}),
$1/3$ (\textit{$\mathrm{C}_{2}A$}), $1/6$ (\textit{$\mathrm{C}_{5}A$})
are shown in Fig. \ref{Fig_Superstructures_substitutional}. Using
the static-concentration-waves' method and the self-consistent field
(mean-field) approximation \cite{Khachaturyan}, one can derive expressions
for the configurational free energies of different honeycomb-lattice-based
structures, 
\begin{equation}
F=U-TS,\label{Eq_free_energy}
\end{equation}
where $U$ and $S$ denote configurational internal energy and entropy,
respectively, and $T$ is an absolute temperature.

Specific (per site) configuration-dependent part of the free energies
for C\textit{A}-type substitutional (super)structures in Figs. \ref{Fig_Superstructures_substitutional}(a)--(c)
are as follows: 
\begin{equation}
F_{1}^{\mathrm{C}A}\cong\frac{1}{2}c^{2}\lambda_{1}(\mathbf{0})+\frac{1}{8}(\eta_{1}^{I})^{2}\lambda_{1}(\mathbf{k}^{M})-TS_{1}^{\mathrm{C}A}(c,\eta_{1}^{I}),\label{Eq_F1_CA}
\end{equation}

\begin{equation}
F_{2}^{\mathrm{C}A}\cong\frac{1}{2}c^{2}\lambda_{1}(\mathbf{0})+\frac{1}{8}(\eta_{2}^{I})^{2}\lambda_{2}(\mathbf{k}^{M})-TS_{2}^{\mathrm{C}A}(c,\eta_{2}^{I}),\label{Eq_F2_CA}
\end{equation}

\begin{equation}
F_{3}^{\mathrm{C}A}\cong\frac{1}{2}c^{2}\lambda_{1}(\mathbf{0})+\frac{1}{8}(\eta_{0}^{I})^{2}\lambda_{2}(\mathbf{0})-TS_{3}^{\mathrm{C}A}(c,\eta_{0}^{I}).\label{Eq_F3_CA}
\end{equation}
Configurational free energies (per site) for $\mathrm{C}_{2}A$-type
substitutional (super)structures presented in Figs. \ref{Fig_Superstructures_substitutional}(h)
and (i) are

\begin{equation}
F_{1}^{\mathrm{C_{2}}A}\cong\frac{1}{2}c^{2}\lambda_{1}(\mathbf{0})+\frac{1}{9}(\eta_{1}^{I})^{2}\lambda_{2}(\mathbf{k}^{K})-TS_{1}^{\mathrm{C_{2}}A}(c,\eta_{1}^{I}).\label{Eq_F1_C2A}
\end{equation}

\begin{equation}
\begin{array}{c}
F_{3}^{\mathrm{C_{2}}A}\cong\frac{1}{2}c^{2}\lambda_{1}(\mathbf{0})+\frac{1}{18}(\eta_{0}^{III})^{2}\lambda_{2}(\mathbf{0})+\frac{1}{36}\left[(\eta_{1}^{III})^{2}+(\eta_{2}^{III})^{2}\right]\lambda_{2}(\mathbf{k}^{K})-\\
-TS_{3}^{\mathrm{C_{2}}A}(c,\eta_{0}^{III},\eta_{1}^{III},\eta_{2}^{III}).
\end{array}\label{Eq_F3_C2A}
\end{equation}
Configurational free energies (per site) for \textit{$\mathrm{C}_{3}A$}-type
substitutional (super)structures presented in Figs. \ref{Fig_Superstructures_substitutional}(d)--(f)
are

\begin{equation}
F_{1}^{\mathrm{C_{3}}A}\cong\frac{1}{2}c^{2}\lambda_{1}(\mathbf{0})+\frac{3}{32}(\eta_{2}^{I})^{2}\lambda_{2}(\mathbf{k}^{M})-TS_{1}^{\mathrm{C_{3}}A}(c,\eta_{2}^{I}),\label{Eq_F1_C3A}
\end{equation}

{\small 
\begin{equation}
F_{2}^{\mathrm{C_{3}}A}\cong\frac{1}{2}c^{2}\lambda_{1}(\mathbf{0})+\frac{1}{32}\left[2(\eta_{1}^{II})^{2}\lambda_{1}(\mathbf{k}^{M})+(\eta_{2}^{II})^{2}\lambda_{2}(\mathbf{k}^{M})\right]-TS_{2}^{\mathrm{C_{3}}A}(c,\eta_{1}^{II},\eta_{2}^{II}),\label{Eq_F2_C3A}
\end{equation}
}{\small \par}

\begin{equation}
\begin{array}{c}
F_{3}^{\mathrm{C_{3}}A}\cong\frac{1}{2}c^{2}\lambda_{1}(\mathbf{0})+\frac{1}{32}\left[(\eta_{0}^{III})^{2}\lambda_{2}(\mathbf{0})+(\eta_{1}^{III})^{2}\lambda_{1}(\mathbf{k}^{M})+(\eta_{2}^{III})^{2}\lambda_{2}(\mathbf{k}^{M})\right]-\\
-TS_{3}^{\mathrm{C_{3}}A}(c,\eta_{0}^{III},\eta_{1}^{III},\eta_{2}^{III}).
\end{array}\label{Eq_F3_C3A}
\end{equation}
Configurational free energy (per site) for $\mathrm{C}_{5}A$-type
substitutional (super)structure {[}Fig. \ref{Fig_Superstructures_substitutional}(j){]}
reads as

\begin{equation}
\begin{array}{c}
F^{\mathrm{C_{5}}A}\cong\frac{1}{2}c^{2}\lambda_{1}(\mathbf{0})+\frac{1}{72}(\eta_{0}^{III})^{2}\lambda_{2}(\mathbf{0})+\frac{1}{36}\left[(\eta_{1}^{III})^{2}+(\eta_{2}^{III})^{2}\right]\lambda_{2}(\mathbf{k}^{K})-\\
-TS^{\mathrm{C_{5}}A}(c,\eta_{0}^{III},\eta_{1}^{III},\eta_{2}^{III}).
\end{array}\label{Eq_F_C5A}
\end{equation}
At last, configurational free energy (per site) for \textit{$\mathrm{C}_{7}A$}-type
substitutional (super)structure {[}Fig. \ref{Fig_Superstructures_substitutional}(g){]}:

{\small 
\begin{equation}
\begin{array}{c}
F^{\mathrm{C_{7}}A}\cong\frac{1}{2}c^{2}\lambda_{1}(\mathbf{0})+\frac{1}{128}\left[(\eta_{0}^{III})^{2}\lambda_{2}(\mathbf{0})+3(\eta_{1}^{III})^{2}\lambda_{1}(\mathbf{k}^{M})+3(\eta_{2}^{III})^{2}\lambda_{2}(\mathbf{k}^{M})\right]-\\
-TS^{\mathrm{C_{7}}A}(c,\eta_{0}^{III},\eta_{1}^{III},\eta_{2}^{III}).
\end{array}\label{Eq_F_C7A}
\end{equation}
} Here, in Eqs. (\ref{Eq_F1_CA})--(\ref{Eq_F_C7A}), $c$ is an atomic
fraction of dopant atoms ($A$), $\eta_{\varsigma}^{\aleph}$ ($\varsigma=0,$
$1$ or $2$) are the long-range order (LRO) parameters ($\Xi$ index
denotes their total number for a given structure; $\aleph=I,$ $II$
or $III$), $\mathbf{k}$ is a wave vector belonging to the first
Brillouin zone of the honeycomb-lattice reciprocal space {[}see Fig.
\ref{Fig_Lattice}(b){]} and generating certain superstructure {[}in
detail see Refs. \cite{Radchenko_NNN-2008,Radchenko_MFiNT,Radchenko_SSP,Radchenko_PhysE}{]}.
Al the thermodynamics of the honeycomb-lattice-based superstructures
in Fig. \ref{Fig_Superstructures_substitutional} is defined by interatomic-interaction
parameters $\lambda_{1}(\mathbf{0})$, $\lambda_{2}(\mathbf{0})$,
$\lambda_{1}(\mathbf{k}^{M})$, $\lambda_{2}(\mathbf{k}^{M})$, $\lambda_{2}(\mathbf{k}^{K})$,
which are connected with pairwise interaction energies $W_{pq}^{\mathrm{CC}}(\mathbf{R}-\mathbf{R}')$,
$W_{pq}^{AA}(\mathbf{R}-\mathbf{R}')$, $W_{pq}^{\mathrm{C\mathit{A}}}(\mathbf{R}-\mathbf{R}')$
between $\mathrm{C}\lyxmathsym{\textendash}\mathrm{C}$, $A\lyxmathsym{\textendash}A$,
$\mathrm{C}\lyxmathsym{\textendash}A$ atoms situated at the sites
of $p$-th and $q$-th ($p$, $q$ $=$ $1$, $2$) sublattices within
the unit sells with origins (``zero'' sites) at $\mathbf{R}$ and
$\mathbf{R}'$. Pairwise interaction energies define so-called mixing
energy, $w_{pq}(\mathbf{R}-\mathbf{R}')\equiv W_{pq}^{\mathrm{CC}}(\mathbf{R}-\mathbf{R}')+W_{pq}^{AA}(\mathbf{R}-\mathbf{R}')-2W_{pq}^{\mathrm{C\mathit{A}}}(\mathbf{R}-\mathbf{R}'),$
which in the literature is known also as ``ordering energy\textquotedblright{}
and ``interchange energy\textquotedblright{} \cite{Khachaturyan,Radchenko_SSP,Radchenko_PhysE}.
The first-, second-, ..., $n$-th-neighbor mixing energies, $w_{1}$,
$w_{2}$, ..., $w_{n}$ {[}see Fig. \ref{Fig_Lattice}(a){]}, are
commonly used for the analysis of the equilibrium atomic order \cite{Radchenko_MFiNT,Radchenko_PhysE,Radchenko_IJHE}
as well as the ordering kinetics \cite{Radchenko_NNN-2008,Radchenko_SSP,Radchenko_SSS}.
For the statistical-thermodynamic description of the interatomic interactions
in all coordination shells, or arbitrary-range interactions, it is
conveniently to apply the Fourier transform, which results to the
interatomic-interaction parameters $\lambda(\mathbf{k})$ entering
into the expressions for the configuration free energies (\ref{Eq_F1_CA})--(\ref{Eq_F_C7A}).

\subsection*{Interstitial Superlattices}

\begin{figure*}[!t]
\includegraphics[width=1\textwidth]{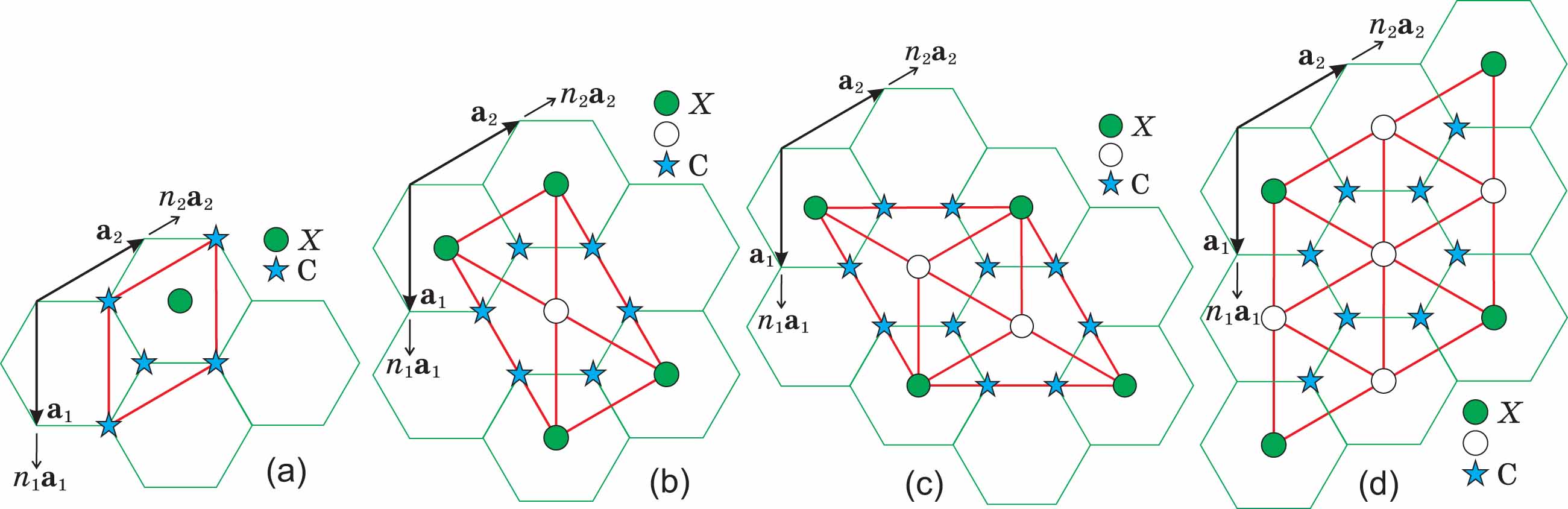}

\caption{(Color online) Primitive unit cells of interstitial superstructures
with stoichiometries $1/3$ (a), $1/5$ (b), $1/7$ (c), $1/9$ (d).
Stars denote carbon atoms, and open circles---unoccupied (in the superstructure)
interstices. }

\label{Fig_Superstructures_interstitial} 
\end{figure*}

Denote interstitial atoms in graphene lattice as $X$, and remained
vacant positions for these atoms in the interstices as $\varnothing$.
Each primitive unit cell of the honeycomb lattice contains two sites
and one interstice being center of the comb {[}Fig. \ref{Fig_Lattice}(a){]}.
An occupation of all interstices by the dopant atoms $X$ corresponds
to the relative impurity concentration $\kappa=\kappa_{st}=1$ and
results to the superstructure-cluster $\mathrm{C}_{2}X$ with a maximal
atomic fraction of the interstitial dopant atoms, $c=c_{st}=1/3$.
Its primitive unit cell is shown in Fig. \ref{Fig_Superstructures_interstitial}(a).
In this case, applying the static concentration waves' approach and
the self-consistent field (mean-field) approximation \cite{Khachaturyan},
one get distribution function for impurity atoms, $P(\mathbf{R})\equiv1$,
and specific configurational free energy (i.e. energy per one interstice)
$F^{\mathrm{C}_{2}X}\cong w(\mathbf{0})/2$. Here $w(\mathbf{k=0})$
is a Fourier-transform of the mixing energy of $X$ and $\varnothing$
components of interstitial subsystem,$ $ $w(\mathbf{R}-\mathbf{R}')\equiv W^{\mathrm{\mathit{XX}}}(\mathbf{R}-\mathbf{R}')+W^{\textrm{\ensuremath{\varnothing\varnothing}}}(\mathbf{R}-\mathbf{R}')-2W^{X\varnothing}(\mathbf{R}-\mathbf{R}')$,
where $W^{\alpha\beta}(\mathbf{R}-\mathbf{R}')$ is energy of effectively
pairwise interaction of $\alpha$ and $\beta$ ($\alpha,\beta=X,\varnothing$)
kinds of ``atoms'' occupying interstices within the primitive unit
cells with radii-vectors $\mathbf{R}$ and $\mathbf{R}'$, respectively.

Specific (per interstice) configurational free energy of $\mathrm{C}_{4}X$-type
interstitial (super)structure {[}Fig. \ref{Fig_Superstructures_interstitial}(b){]},
where in totally ordered state (at 0 K) the relative concentration
of interstitial atoms $\kappa_{st}=1/2$, i.e. their atomic fraction
$c_{st}=1/5$, reads as 
\begin{equation}
F^{\mathrm{C_{4}}X}\cong\frac{1}{2}\kappa^{2}\widetilde{w}(\mathbf{0})+\frac{1}{8}\eta{}^{2}\widetilde{w}(\mathbf{k^{\mathrm{\mathit{M}}}})-TS^{\mathrm{C_{4}}X}(\kappa,\eta).\label{Eq_F_C4X}
\end{equation}
Configurational free energy (per interstice) of $\mathrm{C}_{6}X$-type
interstitial (super)structure in Fig. \ref{Fig_Superstructures_interstitial}(c)
(in the totally ordered state $\kappa_{st}=1/3$, $c_{st}=1/7$) is

\begin{equation}
F^{\mathrm{C_{6}}X}\cong\frac{1}{2}\kappa^{2}\widetilde{w}(\mathbf{0})+\frac{1}{9}\eta{}^{2}\widetilde{w}(\mathbf{k^{\mathrm{\mathit{K}}}})-TS^{\mathrm{C_{6}}X}(\kappa,\eta).\label{Eq_F_C6X}
\end{equation}
In the totally ordered state of $\mathrm{C}_{8}X$-type interstitial
(super)structure {[}Fig. \ref{Fig_Superstructures_interstitial}(d){]},
$\kappa_{st}=1/4$ and $c_{st}=1/9$. Its specific configuration-dependent
part of the free energy reads as

\begin{equation}
F^{\mathrm{C_{8}}X}\cong\frac{1}{2}\kappa^{2}\widetilde{w}(\mathbf{0})+\frac{3}{32}\eta{}^{2}\widetilde{w}(\mathbf{k^{\mathrm{\mathit{M}}}})-TS^{\mathrm{C_{8}}X}(\kappa,\eta).\label{Eq_F_C8X}
\end{equation}

In conclusion of this section, note that all free-energy equations
(\ref{Eq_F1_CA})--(\ref{Eq_F_C8X}), being derived within the framework
of the self-consistent field approximation, are ``governed'' by
the effective pairwise interactions of $\alpha$--$\beta$ particles
only, where $\alpha,\beta=\mathrm{C},A$ for substitutional systems,
and $\alpha,\beta=X,\varnothing$ for interstitial ones. The main
point of such model is that total internal field acting on the substitutional
(interstitial) ``atom'' from the other substitutional (interstitial)
and matrix atoms, is replaced with self-averaged (self-consistent)
field representing the most probable result of the total interaction
of all atoms with distinguished one for a given their distribution
generated by the same (self-consistent) field \cite{Khachaturyan,Bugayev_Tat,Radchenko_NNN-2010}.

\section*{Low-Temperature Stability of Superstructures}

\begin{figure*}
\includegraphics[width=1\textwidth]{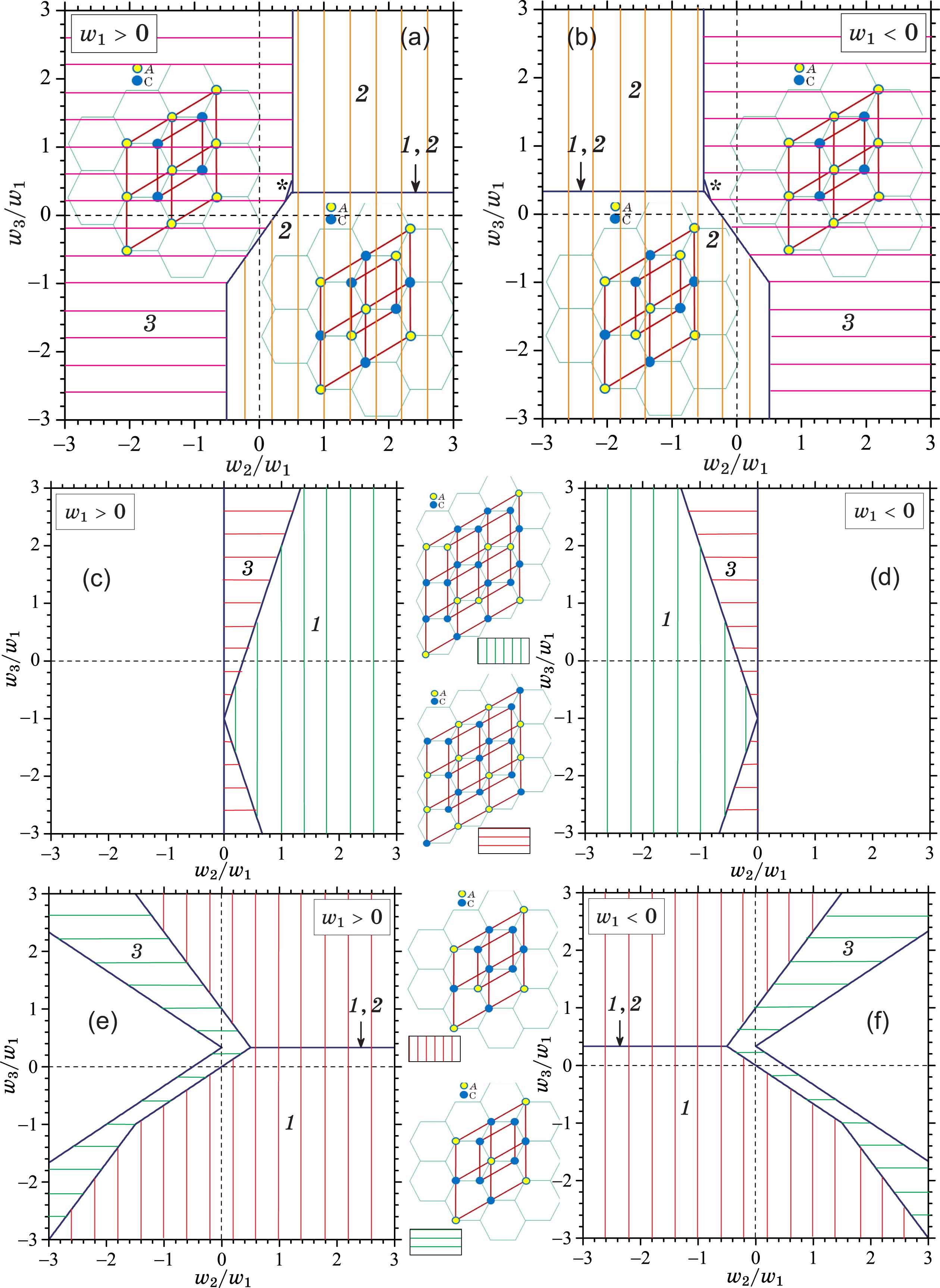}

\caption{(Color online) The low-temperature stability regions (in terms of
the ratios of the mixing energies $w_{2}/w_{1}$ and $w_{3}/w_{1}$)
for $\mathrm{C}A$ (a), (b); $\mathrm{C}_{2}A$ (c), (d); $\mathrm{C_{3}}A$
(e), (f) superstructures assuming interatomic interactions in the
first three coordination shells. Here, (a)--(b) \textit{1}, \textit{2},
\textit{3} denote $\lambda_{1}(\mathbf{k}^{M})$, $\lambda_{2}(\mathbf{k}^{M})$,
$\lambda_{2}(\mathbf{0})$ entering into Eqs. (\ref{Eq_F1_CA})--(\ref{Eq_F3_CA})
for $\mathrm{C}A$; (c)--(f) \textit{1}, \textit{2}, \textit{3} denote
number of LRO parameters describing $\mathrm{C}_{2}A$ and $\mathrm{C_{3}}A$. }

\label{Fig_stability1} 
\end{figure*}

\begin{figure*}[!t]
\includegraphics[width=1\textwidth]{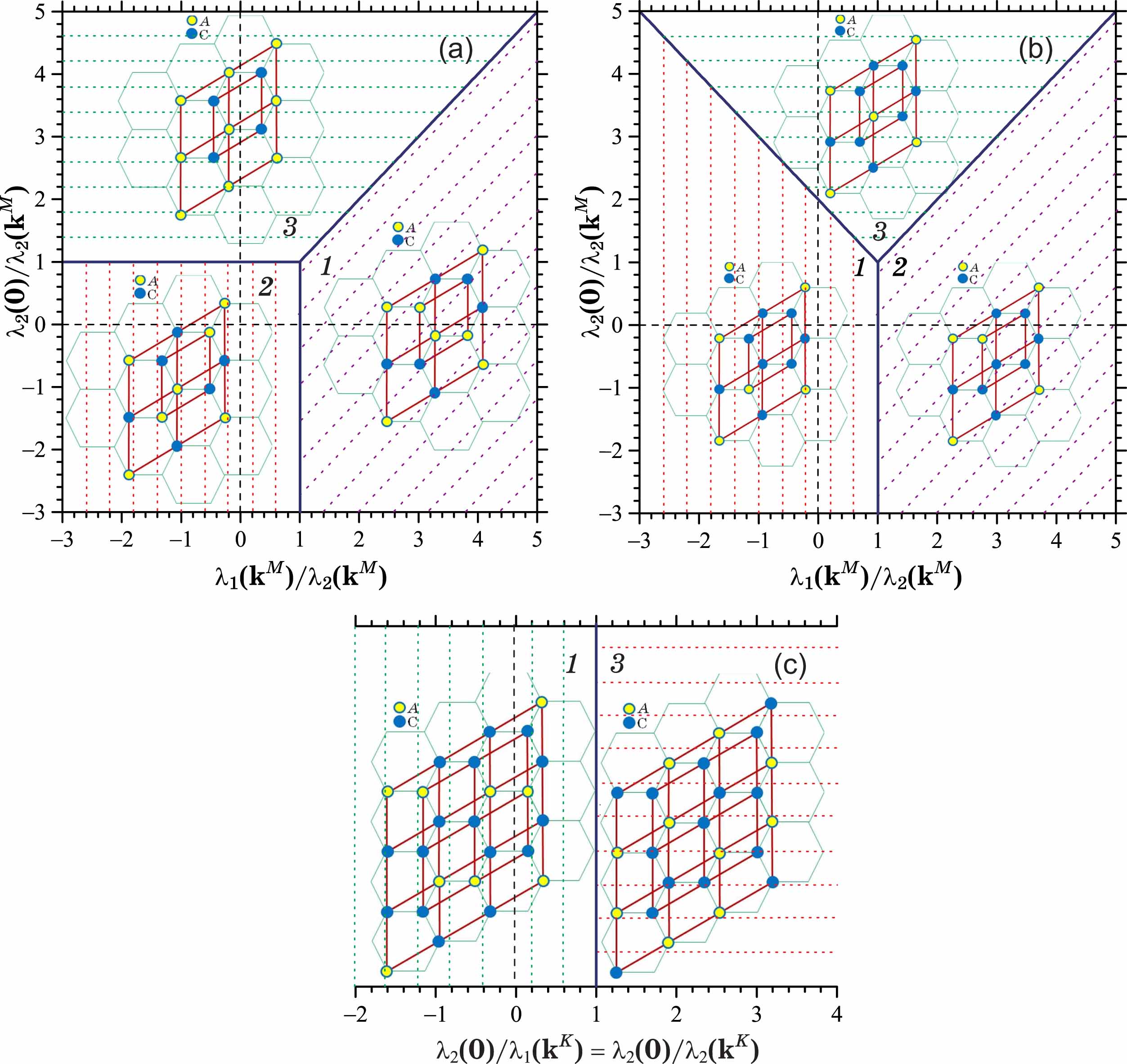}

\caption{(Color online) The same as in the previous figure, but taking into
account interactions of all atoms in the superstructures $\mathrm{C}A$
(a), $\mathrm{C}_{3}A$ (b), and $\mathrm{C_{2}}A$ (c).}

\label{Fig_stability2} 
\end{figure*}

Since each of the stoichiometries among the interstitial superstructures
predicts only one ordered distribution of interstitial atoms, as Figs.
\ref{Fig_Superstructures_interstitial}(a)--(d) demonstrate, below
we pay an attention to the substitutional (super)structural stability
only. (Peculiarities of the stability of interstitial graphene-based
structures can be found in Ref. \cite{Radchenko_NNN-2010}.)

As follows from Eq. (\ref{Eq_free_energy}), the low-temperature (i.e.
at $T=0$ K) stability of a structure, when contribution of the entropy,
$S$, to the total free energy, $F$, is small, depends on the internal
energy, $U$. At $T=0$ K, the stable is a phase which has the lowest
internal energy as compared with other phases of the same composition
(here we are neglecting the possibility of the formation of mechanical
mixture of pure components and different structures). So, to calculate
the low-temperature stability ranges, we minimize the configurational
free energy, $F=U|_{T=0}$, setting in Eqs. (\ref{Eq_F1_CA})--(\ref{Eq_F_C7A})
$T=0$. Such minimization is a sufficient stability condition. The
necessary condition any superstructure to be appeared is a positive
temperature of the stability loss of disordered state with respect
to appearance of the long-range atomic order: $T_{s}=-k_{B}^{-1}c(1-c)\lambda_{\omega}(\mathbf{k})>0$,
i.e., first of all, $\mathrm{min}\lambda_{\omega}(\mathbf{k})<0$
($\omega=1$, $2$; $k\in BZ$) \cite{Khachaturyan}. These two (necessary
and sufficient) conditions can be realized in a certain range of interatomic-interaction
parameters $\lambda(\mathbf{k})$ entering into Eqs. (\ref{Eq_F1_CA})--(\ref{Eq_F_C7A}).

The $\mathrm{C}A$-, $\mathrm{C}_{2}A$-, and $\mathrm{C}_{3}A$-type
superstructures seem the most interesting, since, at these stoichiometries,
there are three or two different (nonequivalent) ordered distributions
of atoms (see Fig. \ref{Fig_Superstructures_substitutional}). The
low-temperature stability regions for these superstructures are represented
in Figs. \ref{Fig_stability1} and \ref{Fig_stability2}, where the
ranges of values of interatomic-interaction parameters providing such
a stability are determined. Two cases are considered: firstly (Fig.
\ref{Fig_stability1}), taking into account only first-, second- and
third-neighbor mixing energies ($w_{1}$, $w_{2}$, $w_{3}$), but
vanishing mixing energies in other (distant) coordination shells,
and, secondly (Fig. \ref{Fig_stability2}), taking into account mixing
energies in all coordination shells.

An account of the third-nearest-neighbor interatomic interactions
always provides the stability for the superstructures {[}Figs. \ref{Fig_Superstructures_substitutional}(c),
(d), (f){]} in which substitutional dopant atoms are surrounded by
the opposite-kind neighbors. However, an account of (only) these (short-range)
interactions can be an inadequate to provide the stability for the
superstructures {[}Figs. \ref{Fig_Superstructures_substitutional}(a)
and (e){]} in which some of the dopant atoms occupy the nearest-neighbor
lattice sites. Figure \ref{Fig_stability2} demonstrates that accounting
of the interactions of all atoms contained in the system yields new
results as compared with those obtained within the scope of the third-nearest-neighbor
interaction approach: every predicted superstructures can be stable
at the appropriate values of interatomic-interaction energies.

At the stoichiometries 1/8 and 1/6, there is only one possible ordered
arrangement of atoms {[}see Figs. \ref{Fig_Superstructures_substitutional}(g),
(j) and also Eqs. (\ref{Eq_F_C5A}), (\ref{Eq_F_C7A}){]}. Therefore,
at low temperatures, $\mathrm{C}_{7}A$- and $\mathrm{C}_{5}A$-type
honeycomb-lattice-based superstructures are stable in all set of interatomic-interaction-energy
values.

Thus, the third-nearest-neighbor Ising model results in the instability
(thermodynamic unfavorableness) of some predicted superstructures.
In contrast to this model, the consideration of all coordination shells
in the interatomic interactions shows that all predicted honeycomb-lattice-based
superstructures are stable at the appropriate values of interatomic-interaction
energies. Moreover, some superstructures {[}$\mathrm{C}A$ and $\mathrm{C}_{3}A$
in Figs. \ref{Fig_Superstructures_substitutional}(a) and (e), respectively{]}
practically may be stable due to the long-range interatomic interactions
only.

The problem of stability for graphene-based structures is considered
at low temperatures. At finite (or room) temperatures, when LRO parameters
in Eqs. (\ref{Eq_F1_CA})--(\ref{Eq_F_C7A}) are not equal to unity,
$\eta_{\varsigma}^{\aleph}\neq1$, an entropy contribution to the
free energy appears. It will result in a shift of the boundaries between
the stability ranges in Figs. \ref{Fig_stability1} and \ref{Fig_stability2},
but it will not change the qualitative results, particularly, the
long-range interatomic-interaction effect on the stability of the
graphene-based (super)structures.

\section*{Kinetics of the Long-Range Atomic-Order Relaxation}

As it is shown above, all interstitial (super)structures (Fig. \ref{Fig_Superstructures_substitutional})
are described by the one LRO parameter only {[}Eqs. (\ref{Eq_F_C4X})--(\ref{Eq_F_C8X}){]},
while this is not the case for substitutional ones (Fig. \ref{Fig_Superstructures_interstitial}),
where two and even there LRO parameters can enter into the free-energy
equations (\ref{Eq_F1_CA})--(\ref{Eq_F_C7A}). That is why here we
consider more complex case---kinetics of the LRO relaxation in the
substitutional systems. (Details of the LRO relaxation in the interstitial
graphene-based systems can be found in Ref. \cite{Radchenko_NNN-2010}.)

Lets us describe the long-range atomic-order kinetics considering
case of exchange (ring) diffusion mechanism governing the atomic ordering
in a two-dimensional binary solid solution $\mathrm{C}_{1-c}A_{c}$
based on a graphene-type lattice (neglecting the vacancies at the
lattice sites). Apply the Önsager-type microdiffusion master equation
\cite{Khachaturyan,Radchenko_SSP,Radchenko_SSS}: 
\begin{equation}
\frac{dP_{p}^{\alpha}(\mathbf{R},t)}{dt}\approx-\frac{1}{k_{B}T}\sum_{\beta=\mathrm{C},A}\,\sum_{q=1}^{2}\,\sum_{\mathbf{R}'}c_{\alpha}c_{\beta}L_{pq}^{\alpha\beta}(\mathbf{R}-\mathbf{R}')\frac{\delta\triangle F}{\delta P_{q}^{\beta}(\mathbf{R}',t)};\label{Eq_Onsager}
\end{equation}
here, $P_{p}^{\alpha}(\mathbf{R},t)$ is a probability to find $\alpha$-atom
in a time $t$ at the $(p,\mathbf{R})$ site, i.e. at the site of
$q$-th sublattice within the unit-cell origin position \textbf{R};
$c_{\alpha}$ ($c_{\beta}$) is a relative fraction of $\alpha$-kind
($\beta$-kind) atom; $||L_{pq}^{\alpha\beta}(\mathbf{R}-\mathbf{R}')||$
is a matrix of the Önsager-type kinetic coefficients whose elements
represent probabilities of elementary exchange-diffusion jumps of
a pair of $\alpha$ and $\beta$ atoms at $\mathbf{r}=\mathbf{R}+\mathbf{h}_{p}$
and $\mathbf{r'}=\mathbf{R'}+\mathbf{h}_{q}$ sites of the $p$-th
and $q$-th sublattices composing the honeycomb lattice and displaced
with respect to each other by the vector \textbf{h} ($\alpha,\beta=\mathrm{C},A$;
$p,q=1,2$; $c_{A}=c$, $c_{\mathrm{C}}=1-c$).

If the vacancy content is small, we have almost identity for the single-site
occupation-probability functions of $A$ and C atoms distribution
over the honeycomb-lattice site: $P_{q}^{\mathrm{C}}(\mathbf{R},t)+P_{q}^{\mathrm{\mathit{A}}}(\mathbf{R},t)\approx1$
$\forall\mathbf{R}\wedge\forall q=1,2\wedge\forall t>0$. Then it
is enough to consider an exchange-microdiffusion migration of only
dopant atoms $A$ in terms of the time dependence of only probabilities
$\{P_{q}(\mathbf{R},t)\}$ $[P_{q}(\mathbf{R},t)\equiv P_{q}^{A}(\mathbf{R},t)\;\forall t>0]$.
One can use kinetics equation (\ref{Eq_Onsager}) to describe microdiffusion
by the other mechanisms since semi-phenomenological Eq. (\ref{Eq_Onsager})
does not contain a certain microdiffusion mechanism. Considering of
any other mechanism does not require changing of the type of Eq. (\ref{Eq_Onsager}),
since diffusion mechanism is defined by the kinetic coefficients $L_{pq}^{\alpha\beta}(\mathbf{R}-\mathbf{R}')$,
which should be linked with microscopic characteristics of the system
(energy barrier heights for atomic jumps, thermal vibrational frequencies
of atoms at the sites, vacancy concentration) and external thermodynamic
parameters (temperature etc.)

Condition of the conservation of each kind of atoms in the system,
assumption that any site is obligatory occupied by C or $A$ atom,
Önsager-type symmetry relations, representation of thermodynamic driven
force $\delta\triangle F/\delta P_{q}(\mathbf{R}')$ (as well as $P_{p}(\mathbf{R})$)
as a superposition of the concentration waves, followed by the Fourier
transform of both members in Eq. (\ref{Eq_Onsager}), yield us differential
equations of the time evolution of the LRO parameters, $\eta_{\varsigma}^{\aleph}$:
\begin{equation}
\frac{d\eta_{\varsigma}^{\aleph}}{dt}\cong-c(1-c)\widetilde{L}(\mathbf{k})\left[\eta_{\varsigma}^{\aleph}\frac{\lambda_{\omega}(\mathbf{k})}{k_{B}T}+\ln Z(c,\eta_{0}^{\aleph},\eta_{1}^{\aleph},\eta_{2}^{\aleph})\right],\label{Eq_LRO-kinetics}
\end{equation}
where $\widetilde{L}(\mathbf{k})$ is the Fourier-component of a concentration-dependent
combination of kinetic coefficients $L_{pq}^{\alpha\beta}(\mathbf{R}-\mathbf{R}')$,
$\widetilde{L}_{pq}^{\alpha\beta}(\mathbf{k})\equiv\sum_{\mathbf{R}}L_{pq}^{\alpha\beta}(\mathbf{R}-\mathbf{R}')\exp[-i\mathbf{k}\cdot(\mathbf{R}-\mathbf{R}')]$,
and particular expressions for $Z(c,\eta_{0}^{\aleph},\eta_{1}^{\aleph},\eta_{2}^{\aleph})$
are presented in Refs. \cite{Radchenko_SSP,Radchenko_SSS}. It is
convenient to solve Eq. (\ref{Eq_LRO-kinetics}) in terms of the reduced
time $t^{*}=\widetilde{L}(\mathbf{k})t$ and temperature $T^{*}=k_{B}T/|\lambda_{\omega}(\mathbf{k})|$.

\begin{figure*}[!t]
\includegraphics[width=1\textwidth]{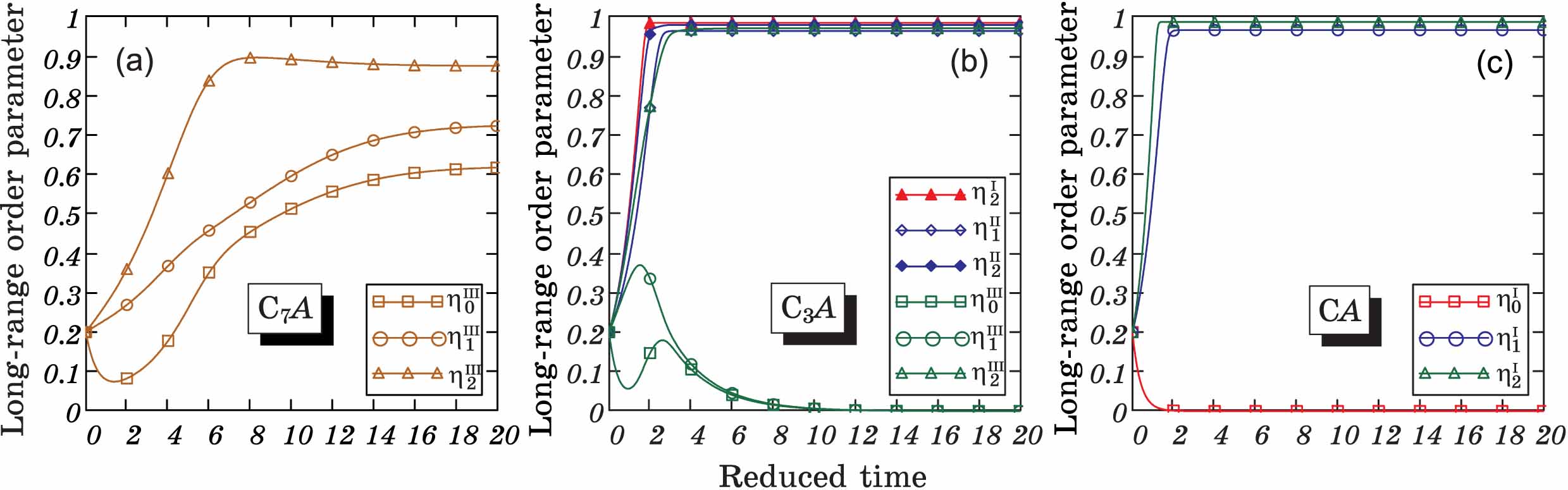}

\caption{(Color online) The time evolution of the LRO parameters in the graphene-based
systems for the temperature $T^{*}=k_{B}T/|\lambda_{2}(\mathbf{k}^{M})|$
and interatomic-interaction parameters parameters $\lambda_{1}(\mathbf{k}^{M})/\lambda_{2}(\mathbf{k}^{M})=5/6$,
$\lambda_{2}(\mathbf{0})/\lambda_{2}(\mathbf{k}^{M})=-5/8$ ($\lambda_{2}(\mathbf{k}^{M})<0$).}

\label{Fig_LRO_kinetics} 
\end{figure*}

Curves in Fig. \ref{Fig_LRO_kinetics} are numerical calculations
of the kinetic equations (\ref{Eq_LRO-kinetics}) for the ordered
$\mathrm{C}_{7}A$, $\mathrm{C}_{3}A$, and $\mathrm{C}A$ superstructural
types at the reduced temperature $T^{*}=0.1$ and certain interatomic-interaction
parameters ${\lambda_{\omega}(\mathbf{k})}$, given as an example.
These values correspond to the certain point {[}$(5/6,-5/8)${]} on
the stability diagrams for $\mathrm{C}A$ and $\mathrm{C}_{3}A$ superstructures
in Figs. \ref{Fig_stability2}(a) and (b). This point indicates what
superstructure is energetically favorable (stable) between the three
possible ones at the given stoichiometry. Stability diagrams in Fig.
\ref{Fig_stability2} are obtained for the absolute zero temperature,
while the kinetic curves in Fig. \ref{Fig_LRO_kinetics} are calculated
for the nonzero temperature. Nevertheless, one can easy see a correspondence
between the statistical-thermodynamic and kinetics results. Results
of the latter improve previous ones; particularly, for the mentioned
point on the diagrams, energetically favorable is a structure described
by the LRO parameter, which relaxes to its equilibrium state being
the highest between the other equilibrium, stationary, and current
values of the LRO parameters of the given composition (see Figs. \ref{Fig_stability2}(a),
(b) and Figs. \ref{Fig_LRO_kinetics}(b), (c).

Figures \ref{Fig_LRO_kinetics}(a) and (b) clear demonstrate that
kinetic curves for the LRO parameters of the $\mathrm{C}_{7}A$- and
$\mathrm{C}_{3}A$-type (super)structures, described by two or three
order parameters, can be nonmonotonic. The nonmonotony is caused not
only by the presence of two interpenetrating sublattices composing
the honeycomb lattice, but also by the dominance of the intersublattice
mixing (interatomic-interaction) energies in their competition with
intrasublattice interaction energies.

\section*{Influence of Correlated and/or Ordered Impurities on Conductivity
of Graphene: Numerical Calculations }

This section is devoted to the investigation of influence of the spatial
correlation and ordering of impurities, acting as a ``disorder''
in graphene, on its conductance using a numerical quantum mechanical
approach. We utilize the time-dependent real-space quantum Kubo--Greenwood
method \cite{Radchenko et al. 1,Radchenko et al. 2,Tuan2013,Roche_SSC,Markussen,Yuan10,Leconte11,Lherbier12,Ishii},
which allows us to study experimentally-relevant large graphene sheets
containing millions of atoms. We consider models of disorder appropriate
for realistic impurities that might exhibit correlations, including
the Gaussian potential describing screened charged impurities and
the short-range potential describing neutral adatoms.

We model electron dynamics in graphene using the standard $p$-orbital
nearest neighbor tight-binding Hamiltonian defined on a honeycomb
lattice \cite{PeresReview,DasSarmaReview}, 
\begin{equation}
\hat{H}=-u\sum_{i,i^{\prime}}c_{i}^{\dagger}c_{i^{\prime}}+\sum_{i}V_{i}c_{i}^{\dagger}c_{i},\label{H}
\end{equation}
where $c_{i}^{\dagger}$ and $c_{i}$ are the standard creation and
annihilation operators acting on a quasi-particle on the site $i$.
The summation over $i$ runs over the entire graphene lattice, while
$i^{\prime}$ is restricted to the sites next to $i$; $u=2.7$ eV
is the hopping integral for the neighboring C atoms $i$ and $i^{\prime}$
with distance $a_{0}\approx0.142$~nm between them (Fig. \ref{Fig_Lattice}),
and $V_{i}$ is the on-site potential describing impurity scattering.

In the present study we consider both short- and long-range impurities.
The short-range impurities represent neutral adatoms on graphene and
are modeled by the delta-function scattering potential for electrons
\begin{equation}
V_{i}=\sum_{j=1}^{N_{imp}}V_{j}\delta_{ij},\label{delta}
\end{equation}
where $N_{imp}$ is the number of impurities on a graphene sheet.
Estimations based on \textit{ab initio} calculations and the \textit{T}-matrix
approach for common adatoms provide typical values for the on-site
potential $V_{j}=V_{0}\lesssim80u$ \cite{Robinson,Wehling,Ihnatsenka,Ferreira},
e.g., for hydrogen, $V_{0}\approx60u$.

The long-range potential is appropriate for screened charged impurities
situated on graphene and/or dielectric substrate. We model them by
the Gaussian scattering potential commonly used in the literature
\cite{PeresReview,DasSarmaReview,Klos} 
\begin{equation}
V_{i}=\sum_{j=1}^{N_{imp}}U_{j}\exp\left(-\frac{|\mathbf{R}_{i}-\mathbf{R}_{j}|^{2}}{2\xi^{2}}\right),\label{Gauss}
\end{equation}
where $\mathbf{R}_{i}$ ($\mathbf{R}_{j}$) is the radius-vector of
the $i$ ($j$) site, $\xi$ is the effective potential radius, and
the potential height is uniformly distributed in the range $U_{j}\in\lbrack-\Delta,\Delta]$
with $\Delta$ being the maximum potential height.

\begin{figure*}
\includegraphics[width=1\textwidth]{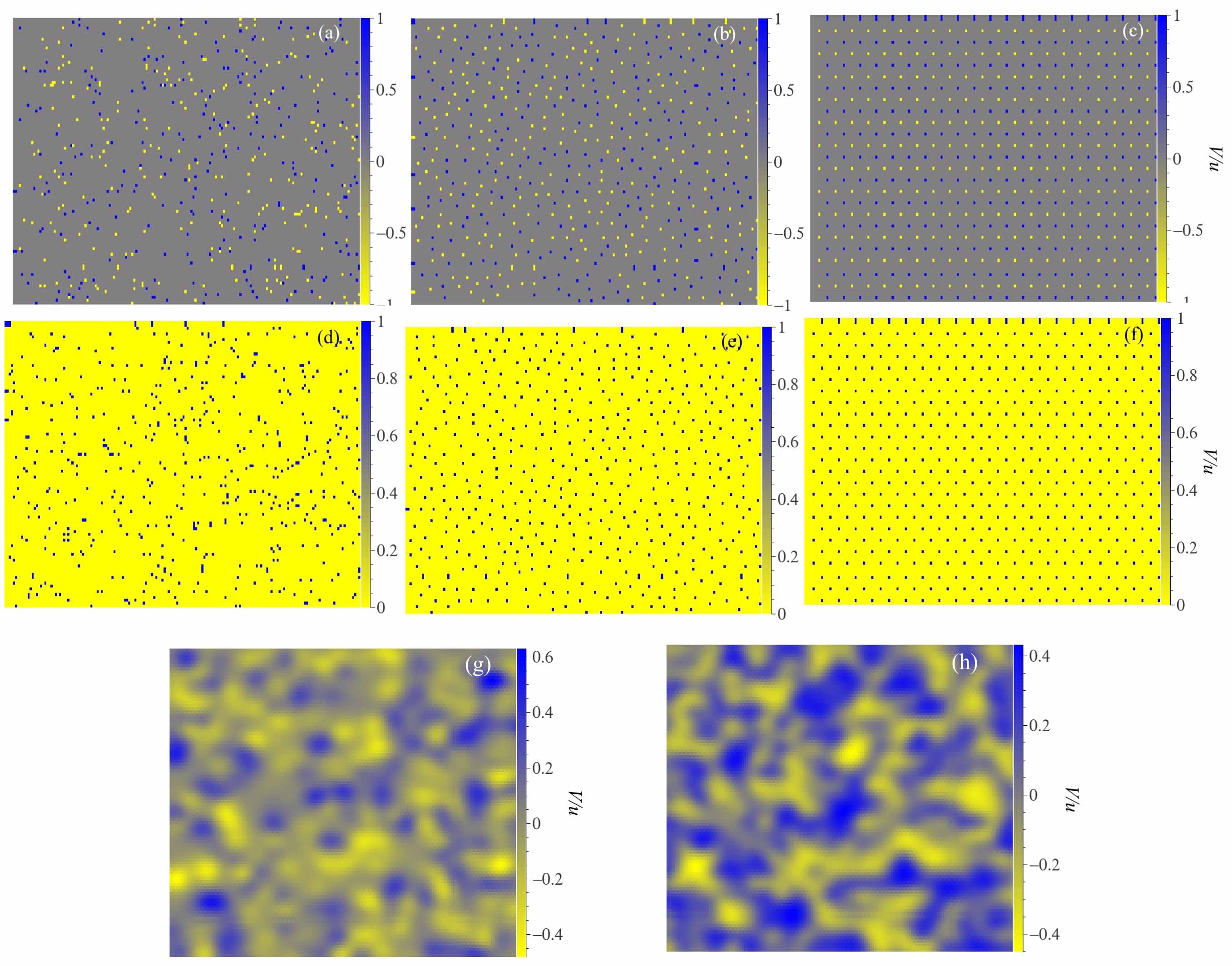}

\caption{(Color online) A representative illustration of (a), (d), (g) random,
correlated (b), (e), and (c), (f), (h) ordered distributions of impurities
for short-range (a)--(c) symmetric (attractive--repulsive) and (d)--(f)
asymmetric (repulsive) scattering potentials, and (g), (d) long-range
Gaussian (attractive--repulsive) potential. }

\label{Fig_Impurity_distribution} 
\end{figure*}

We consider three cases of impurity distribution, random (uncorrelated),
correlated, and ordered. In the first case, the summation in Eqs.
(\ref{delta}), (\ref{Gauss}) is performed over the random distribution
of impurities over the lattice sites. In the second case impurities
are no longer considered to be randomly distributed and to describe
their spatial correlation we adopt a model used in Ref. \cite{Li}
introducing the pair distribution function $P(\mathbf{R}_{i}-\mathbf{R}_{j})\equiv P(r)$,
\begin{equation}
P(r)\equiv\left\{ \begin{array}{c}
0,\; r<r_{0}\text{}\\
1,\; r>r_{0}\text{}
\end{array}\right.\label{g(r)}
\end{equation}
where $r=\mathbf{|R}_{i}-\mathbf{R}_{j}|$ is the distance between
two impurities and the correlation length $r_{0}$ defines the minimum
distance that can separate two impurities. Note, that for the randomly
distributed (totally uncorrelated) impurities $r_{0}=0$. The largest
distance $r_{0_{\max}}$ depends on the relative impurity concentration
$c$: the smaller the concentration $c,$ the larger $r_{0_{\max}}$.
At last, in the third case, the ordered distribution of impurity atoms
over the sites of both sublattices is described by the single-site
occupation-probability functions, which can be derived by the method
of the static concentration waves \cite{Khachaturyan}. Particularly,
for $\mathrm{C}_{7}A$-type ($c_{st}=1/8$) substitutional superstructure
in Fig. \ref{Fig_Superstructures_substitutional}(g), they are \cite{Radchenko_NNN-2008,Radchenko_MFiNT,Radchenko_SSP}
\begin{equation}
\begin{array}{c}
\begin{array}{c}
\left(\begin{array}{c}
P_{1}(\mathbf{R})\\
P_{2}(\mathbf{R)}
\end{array}\right)=c\left(\begin{array}{c}
1\\
1
\end{array}\right)+\frac{1}{8}\eta_{0}^{III}\left(\begin{array}{r}
1\\
-1
\end{array}\right)+\\
+\frac{1}{8}\eta_{1}^{III}\left[\left(\begin{array}{c}
1\\
1
\end{array}\right)\cos\pi n_{1}+\left(\begin{array}{r}
1\\
-1
\end{array}\right)\cos\pi n_{2}+\left(\begin{array}{c}
1\\
1
\end{array}\right)\cos\pi(-n_{1}+n_{2})\right]+
\end{array}\\
\frac{1}{8}\eta_{2}^{III}\left[\left(\begin{array}{r}
1\\
-1
\end{array}\right)\cos\pi n_{1}+\left(\begin{array}{r}
1\\
1
\end{array}\right)\cos\pi n_{2}+\left(\begin{array}{r}
1\\
-1
\end{array}\right)\cos\pi(-n_{1}+n_{2})\right];
\end{array}\label{Eq_P_1/8}
\end{equation}
$n_{1},$ $n_{2},$ $n_{3}$ are integers. $P_{1}(\mathbf{R})$ and
$P_{2}(\mathbf{R})$ possess four values {[}$c+\frac{1}{8}(\eta_{0}^{III}+3\eta_{1}^{III}+3\eta_{2}^{III})$,
$c+\frac{1}{8}(\eta_{0}^{III}-\eta_{1}^{III}-\eta_{2}^{III})$, $c+\frac{1}{8}(-\eta_{0}^{III}+\eta_{1}^{III}-\eta_{2}^{III})$,
$c+\frac{1}{8}(-\eta_{0}^{III}-3\eta_{1}^{III}+3\eta_{2}^{III})${]}
over all lattice sites. The representative examples of random and
correlated distributions for the cases of the short- and long-range
potentials are shown in Fig. \ref{Fig_Impurity_distribution}.

\begin{figure*}
\includegraphics[width=1\textwidth]{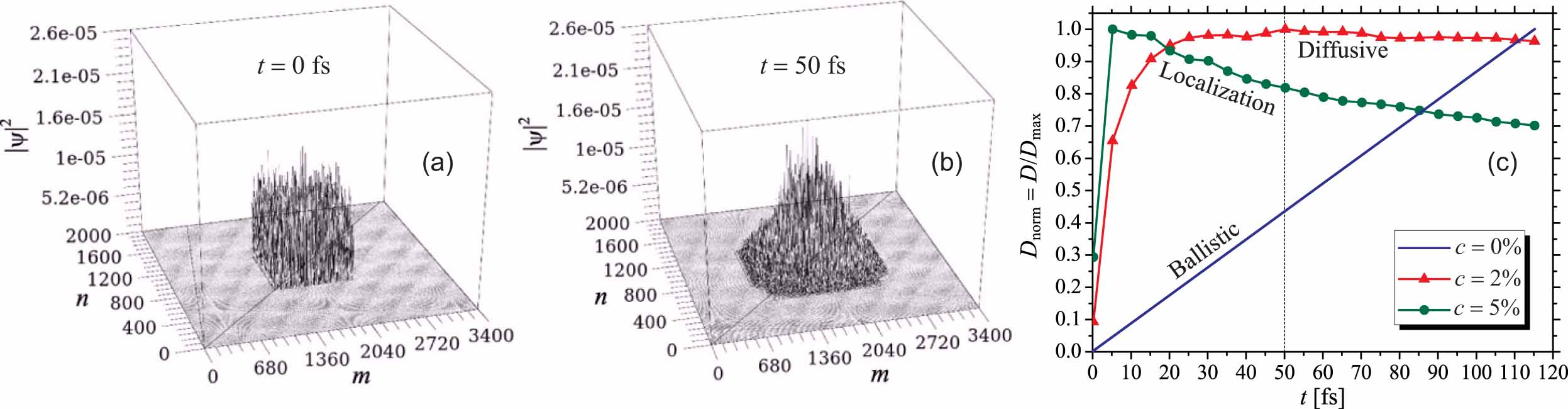}

\caption{(Color online) Wave-packet propagation (a) and (b) in graphene lattice
without ($c=0\%$) and with short-range strong impurities ($c=2\%$
and $5\%$) modeled by the onsite potential $V\sim37u$; temporal
evolution of the $D_{\mathrm{norm}}(E,t)=D(E,t)/D_{\mathrm{max}}(E)$
value normalized by the diffusion coefficient $D_{\mathrm{max}}(E)$
(c). Transport curves are presented for $E=0.2u$.}

\label{Fig_Diffusivity} 
\end{figure*}

The transport properties of graphene sheets can be calculated on the
basis of the time-dependent real-space Kubo formalism, extracting
the dc conductivity $\sigma$ from the wave packet temporal dynamics
governed by the time-dependent Schrödinger equation \cite{Radchenko et al. 1,Radchenko et al. 2,Tuan2013,Roche_SSC,Markussen,Yuan10,Leconte11,Lherbier12,Ishii}.

A central quantity in the Kubo--Greenwood approach is the mean quadratic
spreading of the wave packet along the $x$-direction at the energy
$E$, $\Delta\hat{X}^{2}(E,t)=\bigl\langle\hat{(X}(t)-\hat{X}(0))^{2}\bigr\rangle$,
where $\hat{X}(t)=\hat{U}^{\dagger}(t)\hat{X}\hat{U}(t)$ is the position
operator in the Heisenberg representation, and $\hat{U}(t)=e^{-i\hat{H}t/\hbar}$
is the time-evolution operator {[}wave-packet propagation is visualized
in Figs. \ref{Fig_Diffusivity}(a), (b){]}. Starting from the Kubo--Greenwood
formula for the dc conductivity \cite{Madelung}

\begin{equation}
\sigma=\frac{2\pi\hbar e^{2}}{\Omega}\text{Tr}[\hat{v}_{x}\delta(E-\hat{H})\hat{v}_{x}\delta(E-\hat{H})],\label{Eq_Kubo-Greenwood}
\end{equation}
where $\hat{v}_{x}$ is the $x$-component of the velocity operator,
$E$ is the Fermi energy, $\Omega$ is the area of the graphene sheet,
and factor 2 accounts for the spin degeneracy, the conductivity can
then be expressed as the Einstein relation, 
\begin{equation}
\sigma\equiv\sigma_{xx}=e^{2}\tilde{\rho}(E)\lim_{t\rightarrow\infty}D(E,t),\label{Eq_sigma(t)}
\end{equation}
where $\tilde{\rho}(E)=\frac{\rho}{\Omega}=\frac{\textrm{Tr}[\delta(E-\hat{H})]}{\Omega}$
is the density of electronic sates (DOS) per unit area (per spin),
and the time-dependent transport coefficient $D(E,t)$ (commonly called
as diffusivity) relates to $\Delta\hat{X}^{2}(E,t)$ as 
\begin{align}
D(E,t)=\frac{\bigl\langle\Delta\hat{X}^{2}(E,t)\bigr\rangle}{t}= & \frac{1}{t}\frac{\text{Tr}[\hat{(X}_{H}(t)-\hat{X}(0))^{2}\delta(E-\hat{H})]}{\text{Tr}[\delta(E-\hat{H})]}.\label{Eq_Diffusion}
\end{align}

Further, we are interested in the diffusive transport regime at $t\rightarrow\infty$
{[}Fig. \ref{Fig_Diffusivity}(c){]}, when, neglecting the quantum-localization
effects, the coefficient $D(E,t)$ reaches its maximum. Therefore,
following Refs. \cite{Leconte11,Lherbier12}, we replace in Eq.~(\ref{Eq_sigma(t)})
$\lim_{t\rightarrow\infty}D(E,t)\rightarrow D_{\max}(E),$ such that
the diffusion-controlled dc conductivity is defined as 
\begin{equation}
\sigma=e^{2}\tilde{\rho}(E)D_{\max}(E).\label{Eq_sigmaMax}
\end{equation}
Note that in most experiments, the conductivity is measured as a function
of electron density $n_{e}$. We calculate the electron density as
$n_{e}(E)\equiv n_{e}=\int_{-\infty}^{E}\tilde{\rho}(E)dE-n_{\text{ions}},$
where $n_{\text{ions}}=3.9\cdot10^{15}$ cm$^{-2}$ is the density
of the positive ions in the graphene lattice compensating the negative
charge of the $p$-electrons {[}for the ideal graphene lattice, i.e.
without defects, at the neutrality point $n_{e}(E)=0${]}. Combining
the calculated $n_{e}(E)$ with $\sigma(E)$ given by Eq.~(\ref{Eq_sigmaMax}),
one can obtain the required dependence of the conductivity $\sigma=\sigma(n_{e})$.
Details of numerical calculations of DOS, $D(E,t)$, and $\sigma$
are given in Ref. \cite{Radchenko et al. 1}.

\begin{figure*}[!t]
\includegraphics[width=1\textwidth]{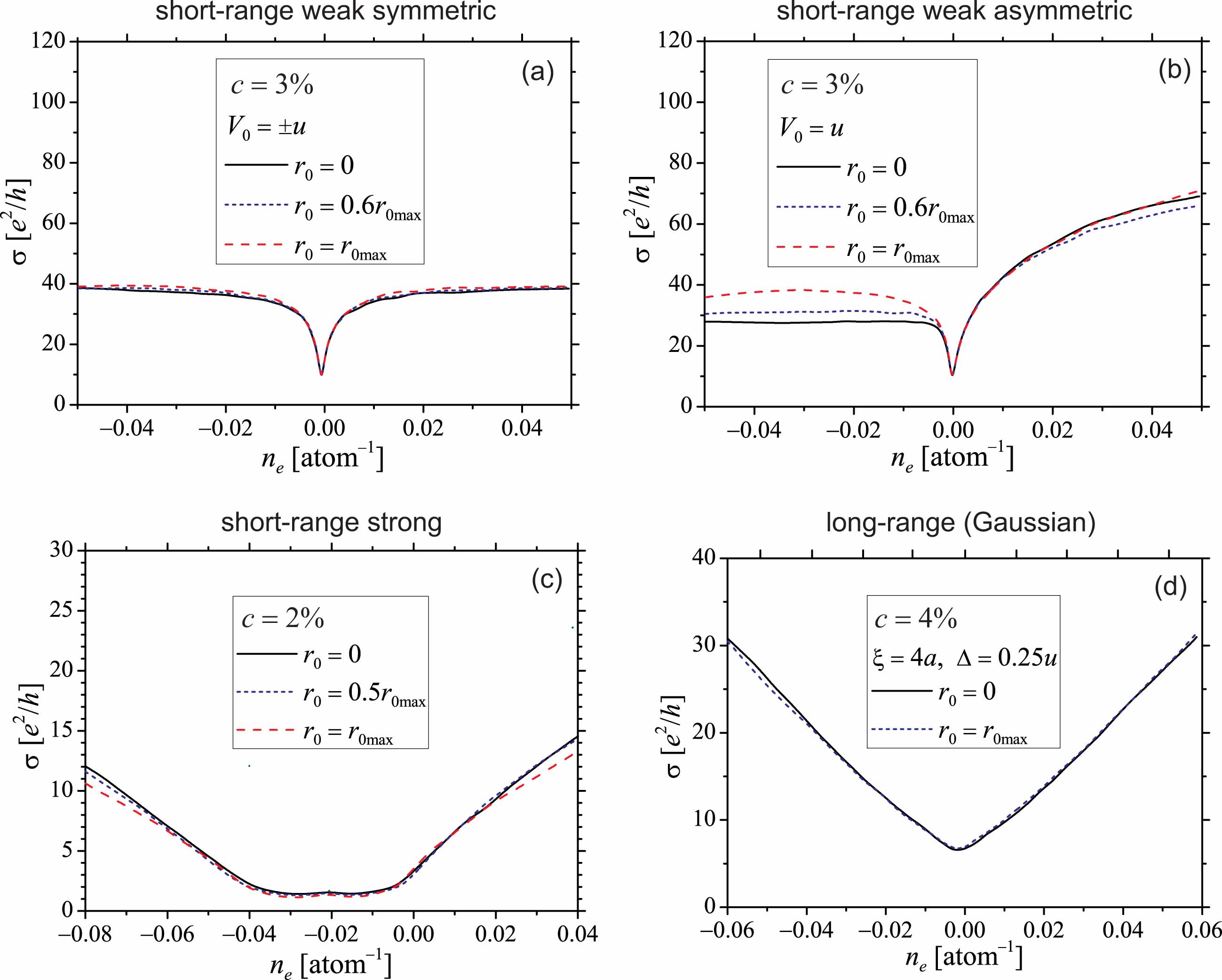}

\caption{(Color online) Conductivity $\sigma$ as a function of the relative
charge carrier (electron) density $n_{e}$ (the number of electrons
per C atoms) for different concentration $c$ of random ($r_{0}=0$)
and correlated ($r_{0}=\frac{1}{2}r_{0_{\max}}$, $r_{0}=r_{0_{\max}}$)
short-range weak (a) and(b), short-range strong (c), and long-range
Gaussian (d) impurities.}

\label{Fig_Cond_corr} 
\end{figure*}

\begin{figure*}[!t]
\includegraphics[width=1\textwidth]{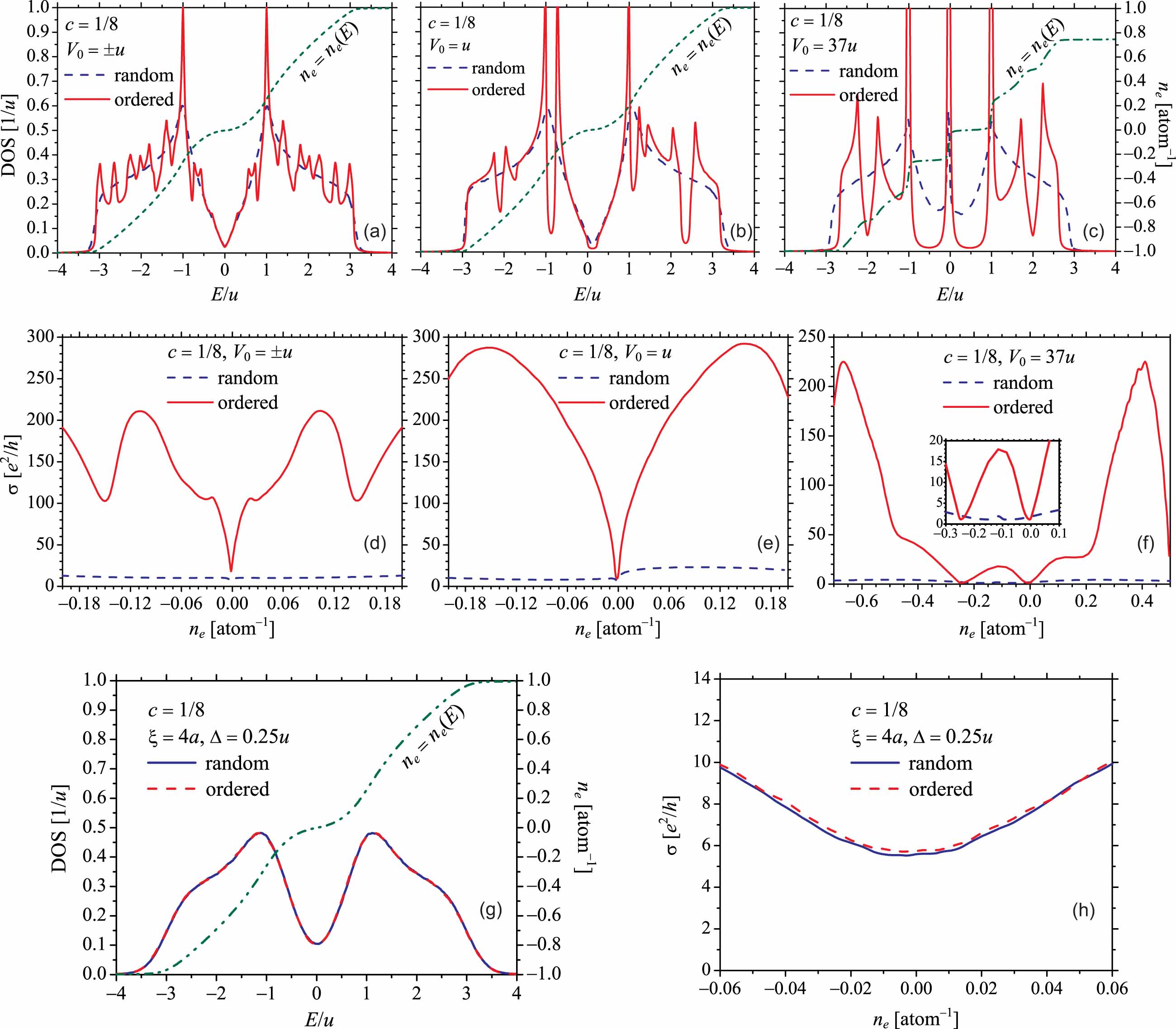}

\caption{(Color online) Density of electronic states (a)--(c), (g) and conductivity
(d)--(f), (h) for 12.5\% of random and ordered impurities, modeled
by the short-range weak symmetric (a) and (d), short-range weak asymmetric
(b) and (e), short-range strong (c) and (f), and long-range Gaussian
(g) and (h) scattering potentials. }

\label{Fig_DOS_Cond_ord} 
\end{figure*}

Figure \ref{Fig_Cond_corr} shows the electron-density dependencies
of the conductivity $\sigma=\sigma(n_{e})$ for random and correlated
impurities modeled by different scattering potentials, where positive
and negative values of $n_{e}$ correspond to different kinds of charge
carriers: electrons and holes. As Figure \ref{Fig_Cond_corr} demonstrates,t
for the most important, experimentally relevant cases of point defects,
namely the strong short-range potential and the long-range Gaussian
potential, the correlation in the distribution of impurity atoms does
not affect the conductivity of the graphene as compared to the case
when they are distributed randomly. This represent the main result
for the case of the correlation. We find that the correlations lead
to the enhancement of the conductivity only for the case of the weak
short-range potential and only when the potential is asymmetric, i.e.
$V=V_{0}$ or $V=-V_{0}.$ No enhancement of the conductivity is found
for the symmetric weak short-range potential, $V=\pm V_{0}.$

As it was mentioned in the introduction, in the recent experiment
\cite{YanFuhrer} the temperature increase of the conductivity was
attributed to the enhancement in the spatial correlation of the adsorbed
potassium ions. Numerical findings do not sustain this interpretation,
the obtained here results strongly suggest that the enhancement of
the conductivity reported in Ref. \cite{YanFuhrer} is most likely
caused by other factors not related to the correlations of impurities.
The numerical calculations do not support also theoretical predictions
in Ref. \cite{Li} that the correlations in the impurity positions
for the long-range potential lead to the enhancement of the conductivity.
This can be attributed to the utilization of the standard Boltzmann
approach within the Born approximation which is not valid for the
case of a long-range potential in the parameter range corresponding
to realistic systems.

In contrast to the case of the correlated impurities, the ordered
short-range weak and essentially strong impurities can strongly affect
charge transport in graphene {[}Fig. \ref{Fig_DOS_Cond_ord}(a)--(f){]}.
In the DOS-curves discrete energy levels appear {[}Figs. \ref{Fig_DOS_Cond_ord}(a)--(c){]}
and broaden as impurity concentration or/and scattering potential
increases. This oscillations (peaks) in the DOS and therefore in the
conductivity are caused by the strongly periodic potential, describing
periodic positions of impurity atoms ``precipitating'' a certain
superstructure (in the given case, with the stoichiometry $c_{st}=1/8$
). It is clear seen from Fig. \ref{Fig_DOS_Cond_ord} that the conductivity
curves for the ordering case rise up to tens times for the short-range
weak {[}Figs. \ref{Fig_DOS_Cond_ord}(d), (e){]} and especially strong
{[}Fig. \ref{Fig_DOS_Cond_ord}(f){]} scatterers as compared with
the case of the randomly-distributed scattering centers. However,
as for the correlation, ordering does not affect nor density of electronic
states nor conductivity for the long-range (Gaussian) scattering potential
(Figs. \ref{Fig_DOS_Cond_ord}(g), (h).

\section*{Anisotropy and Increase in Conductivity due to the Orientational
Correlation of Line Defects in Graphene}

Nowadays, several techniques are capable of producing high-quality,
large-scale graphene. These include CVD-grown graphene on transition
metal surfaces \cite{Kim2009} and epitaxial graphene growth on SiC
\cite{de Heer2007}. Usually, the growth of graphene by the CVD-method
requires to use metal surfaces with hexagonal symmetry, such as the
(111) surface of cubic or the (0001) surface of hexagonal crystals
\cite{Krasheninnikov2011}. The mismatch between the metal-substrate
and graphene causes the strains in the latter, reconstructs the chemical
bonds between the carbon atoms and results in formation of two-dimensional
(2\textit{D}) domains of different crystal orientations separated
by one-dimensional defects \cite{Krasheninnikov2011,Jeong2008,Yazyev2010,Malola2010}.
The nucleation of the graphene phase takes place simultaneously at
different places, which leads to the formation of independent 2\textit{D}
domains matching corresponding grains in the substrate. A line defect
appears when two graphene grains with different orientations coalesce;
the stronger the interaction between graphene and the substrate, the
more energetically preferable the formation of line defects is. These
line defects accommodate localized states trapping the electrons,
originating lines of immobile charges that scatter the Dirac fermions
in graphene. It is well established that the presence of grains and
grain boundaries in three-dimensional polycrystalline materials can
strongly affect their electronic and transport properties. Hence,
in principle, the role of such structures in 2\textit{D} materials,
such as graphene, can be even more important because even a single
line defect can divide and disrupt the crystal \cite{Krasheninnikov2011}.
Theoretical results \cite{Radchenko et al. 2,FerreiraEPL} improve
that the presence of charged line defects strongly affect the transport
properties of CVD-graphene. Such effect becomes more weighty due to
existence of ordered line defects in CVD-synthesized graphene \cite{Ni2012}.

In epitaxial graphene the surface steps caused by substrate morphology
are spatially correlated and act as line scatterers for the charge
carriers \cite{Kuramochi2012}. Epitaxial graphene films grown on
SiC \cite{Kuramochi2012,Held2012} (by SiC decomposition) or on Ru
\cite{Gunther2011} (by CVD method) comprise two distinct self-organized
periodic regions of terrace and step, leading to ordered graphene
domains \cite{Gunther2011}. Experimental measurements show an increase
of the resistance with the step density \cite{Dimitrakopoulos2001},
the step heights \cite{Ji2011}, the step bunching \cite{Lin2011}.
Also, an anisotropy of the conductivity in the parallel and perpendicular
directions to the steps is revealed, which is due to to higher defect
abundance in the step regions \cite{Kuramochi2012,Yakes2010}. Substrate
steps alone increase the resistivity in several times relative to
a perfect terrace \cite{Ji2011} with the ratio of the estimated electron
mobilities in the terrace and step regions being about 10:1 \cite{Kuramochi2012}.
Despite the strong curvature of graphene in the vicinity of steps,
a structural deformation contributes only little to electron scattering
\cite{Low2012}. For the SiC substrate, the dominant scattering mechanism
is provided by the sharp potential variations in the vicinity of the
step due to the electrostatic doping from the substrate strongly coupled
with graphene in the step regions \cite{Low2012}.

Several\textbf{ }theoretical studies have been recently reported addressing
transport properties of graphene with a single graphene boundary \cite{YazyevNature,Liwei,Peres}
or polycrystalline graphene with many domain boundaries \cite{Tuan2013}.\textbf{
}On the other hand, much less attention has been paid to the effect
of charge accumulation at these boundaries due to self-doping. Transport
properties of graphene with 1\textit{D} charged defects has been studied
in Ref. \cite{FerreiraEPL} using the Boltzmann approach within the
first Born approximation. It has been demonstrated that such approximation
is not always applicable for the description of electron transport
in graphene even at finite (non-zero) electronic densities \cite{Ferreira,HengyiXu2011,Klos}.
Following Ref. \cite{Radchenko et al. 2}, below we present results
of investigations of the impact of extended charged defects in the
transport properties of graphene by an exact numerical approach based
on the time-dependent real-space quantum Kubo method \cite{Radchenko et al. 1,Radchenko et al. 2,Tuan2013,Roche_SSC,Markussen,Yuan10,Leconte11,Lherbier12,Ishii},
which is especially suited for experimentally-relevant systems containing
millions of atoms.

\begin{figure*}
\includegraphics[width=1\textwidth]{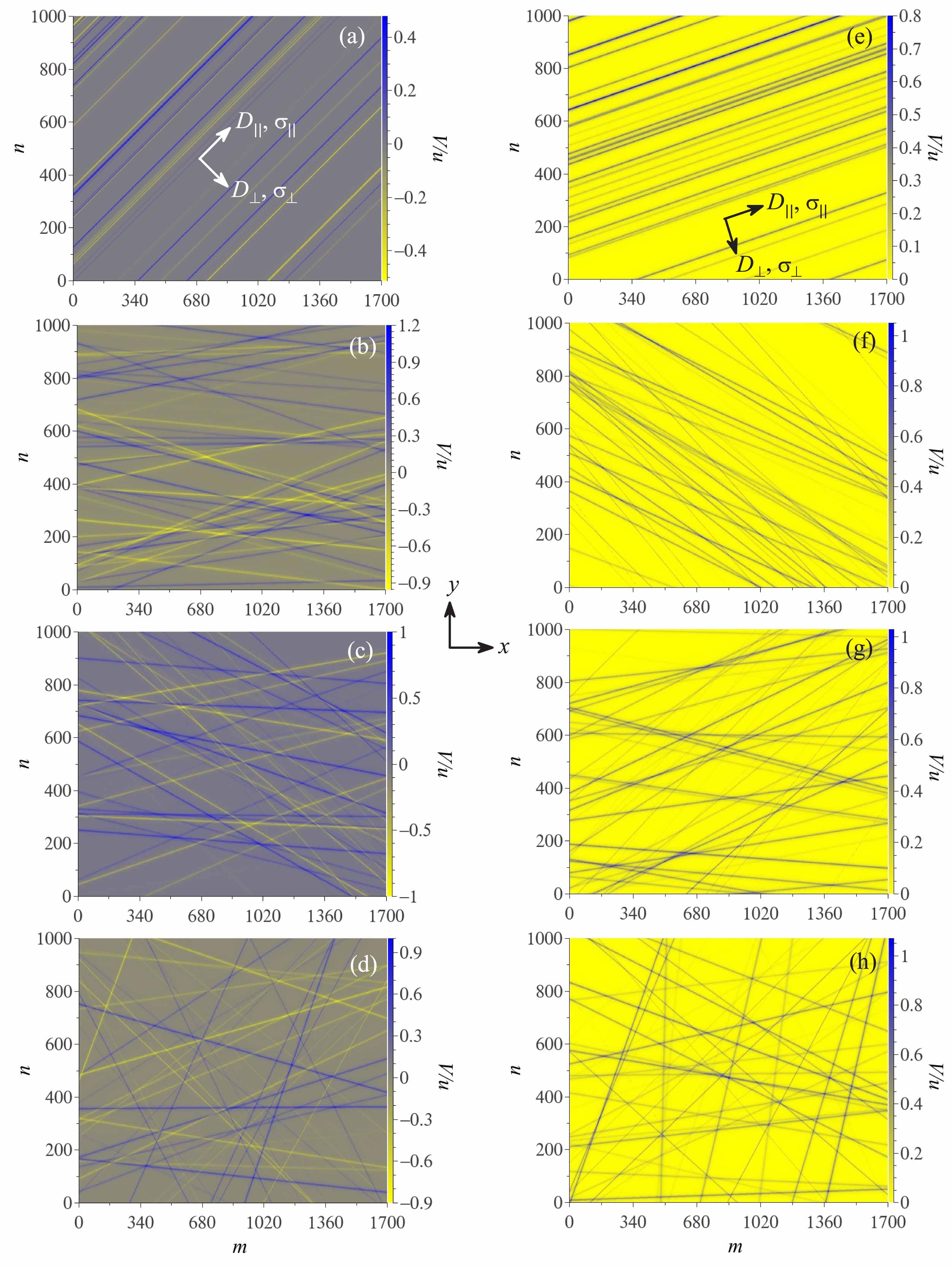}

\caption{(Color online) Effective symmetric (attractive--repulsive), $V\gtrless0$,
(a)--(d) and asymmetric (repulsive), $V>0$, (e)--(h) scattering potentials
for a representative configuration of 50 orientationally-correlated
line defects with different correlation angles $\alpha_{\textrm{max}}$
(the maximal possible angle between any two lines): $0^{\circ}$ (a),
(e); $30^{\circ}$ (b), (f); $60^{\circ}$ (c), (g); $90^{\circ}$
(d), (h). Note, that $\alpha_{\textrm{max}}=0^{\circ}$ and $\alpha_{\textrm{max}}=90^{\circ}$
correspond to the cases of parallel and random (totally uncorrelated)
lines, respectively. Maximum potential height $\triangle=0.25u$. }

\label{Fig_Line-potential} 
\end{figure*}

Since line defects can be thought as lines of reconstructed point
defects \cite{Jeong2008,Krasheninnikov2011,Yazyev2010,Malola2010},
we model a 1\textit{D} defect as point defects oriented along a fixed
direction (corresponding to the line direction) in the honeycomb lattice.
The electronic effective potential for a charged line within the Thomas--Fermi
approximation was first obtained in Ref. \cite{FerreiraEPL}. If there
are $N_{\text{lines}}$ such charged lines in a graphene lattice,
the effective scattering potential reads as 
\begin{equation}
V_{i}=\sum_{j=1}^{N_{\text{lines}}}U_{j}\left[-\text{cos}(q_{\mathrm{\textrm{TF}}}x_{ij})\text{Ci}(q_{\textrm{TF}}x_{ij})+\text{sin}(q_{\textrm{TF}}x_{ij})\left(\frac{\pi}{2}-\text{Si}(q_{\textrm{TF}}x_{ij})\right)\right],\label{Eq_TF potential}
\end{equation}
where $U_{j}$ is a potential height, $x_{ij}$ is a distance between
the site $i$ and the $j$-th line, $q_{\textrm{TF}}=e^{2}k_{F}/(\pi\varepsilon_{0}\varepsilon_{\textrm{r}}\hbar v_{F})$
is the Thomas--Fermi wave-vector defined by the electron Fermi velocity
$v_{F}=3ua/(2\hbar)$ and the Fermi momentum $k_{F}=\sqrt{\pi|n_{e}|}$
(related to the electronic carrier density $n_{e}$ controlled applying
the back-gate voltage). Here, $-e<0$ denotes the electron charge.
The Thomas--Fermi wave-vector is also commonly expressed as a function
of graphene's structure constant $\alpha_{\textrm{g}}\equiv e^{2}k_{F}/(4\pi\varepsilon_{0}\varepsilon_{\textrm{r}}\hbar v_{F})$
according to $q_{\textrm{TF}}=4\alpha_{\textrm{g}}k_{F}$. We consider
two cases: symmetric (attractive and repulsive), $V\gtrless0$, and
asymmetric (repulsive), $V>0$, potentials, where $U_{j}$ are chosen
randomly in the ranges $\left[-\triangle,\triangle\right]$ and $\left[0,\triangle\right]$,
respectively, with $\triangle$ being the maximal potential height.
In order to simplify numerical calculations, we fit the potential
(\ref{Eq_TF potential}) by the Lorentzian (Cauchy) function 
\begin{equation}
V_{i}=\sum_{j=1}^{N_{\text{lines}}}U_{j}\frac{A}{B+Cx_{ij}^{2}},\label{Eq_Lorentzian}
\end{equation}
where the fitting parameters $A$, $B$, $C$ can be calculated from
the least-squares method \cite{Radchenko et al. 2}. The typical shapes
of the effective potential for both symmetric (attractive--repulsive)
and asymmetric (repulsive) cases are illustrated in Fig. \ref{Fig_Line-potential}.

\begin{figure*}
\includegraphics[width=1\textwidth]{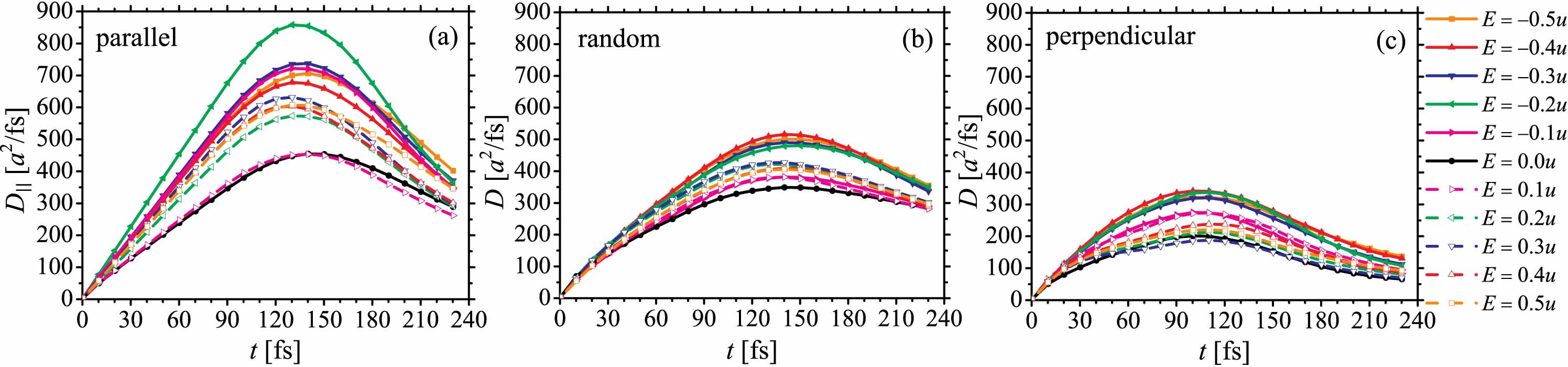}

\caption{(Color online) Time-dependent electron transport coefficients within
the energy interval $E\in[-0.5u,0.5u]$ for 50 parallel (a), randomly
distributed (b), and perpendicular (c) line defects. $D_{\Vert}$
and $D_{\bot}$ denote the transport coefficients in parallel and
perpendicular directions to the lines, $D\equiv D{}_{xx}$ is the
transport coefficient along the \textit{x} direction (see Fig. \ref{Fig_Line-potential}).
The scattering potential is symmetric ($V\gtrless0$), the maximal
potential height $\triangle=0.25u$. }

\label{Fig_Line-diffusivity} 
\end{figure*}

Figure \ref{Fig_Line-diffusivity} shows the time evolution of the
diffusion coefficient within the energy interval $E\in[-0.5u,0.5u]$
for the symmetric (attractive--repulsive) potential for three different
cases of orientation distribution of 50 line defects. (Transport curves
for the case of the asymmetric potential exhibit a similar behavior
and are not shown here). In the first and the third cases, Figs. \ref{Fig_Line-diffusivity}
(a) and (c), the transport coefficients $D_{\Vert}$ and $D_{\bot}$
are calculated respectively along and across 50 parallel-oriented
lines (distance between them is different and random). In the second
case, Fig. \ref{Fig_Line-diffusivity}(b), the lines are randomly
distributed, which results in the isotropic transport, i.e. $D_{{\normalcolor \textrm{rnd}}}\equiv D{}_{xx}\equiv D{}_{yy}$.
As expected, the electron transport along the lines are higher than
those across the lines, whereas $D_{\bot}<D_{\mathrm{rnd}}<D_{\Vert}$.

\begin{figure*}[!t]
\includegraphics[width=1\textwidth]{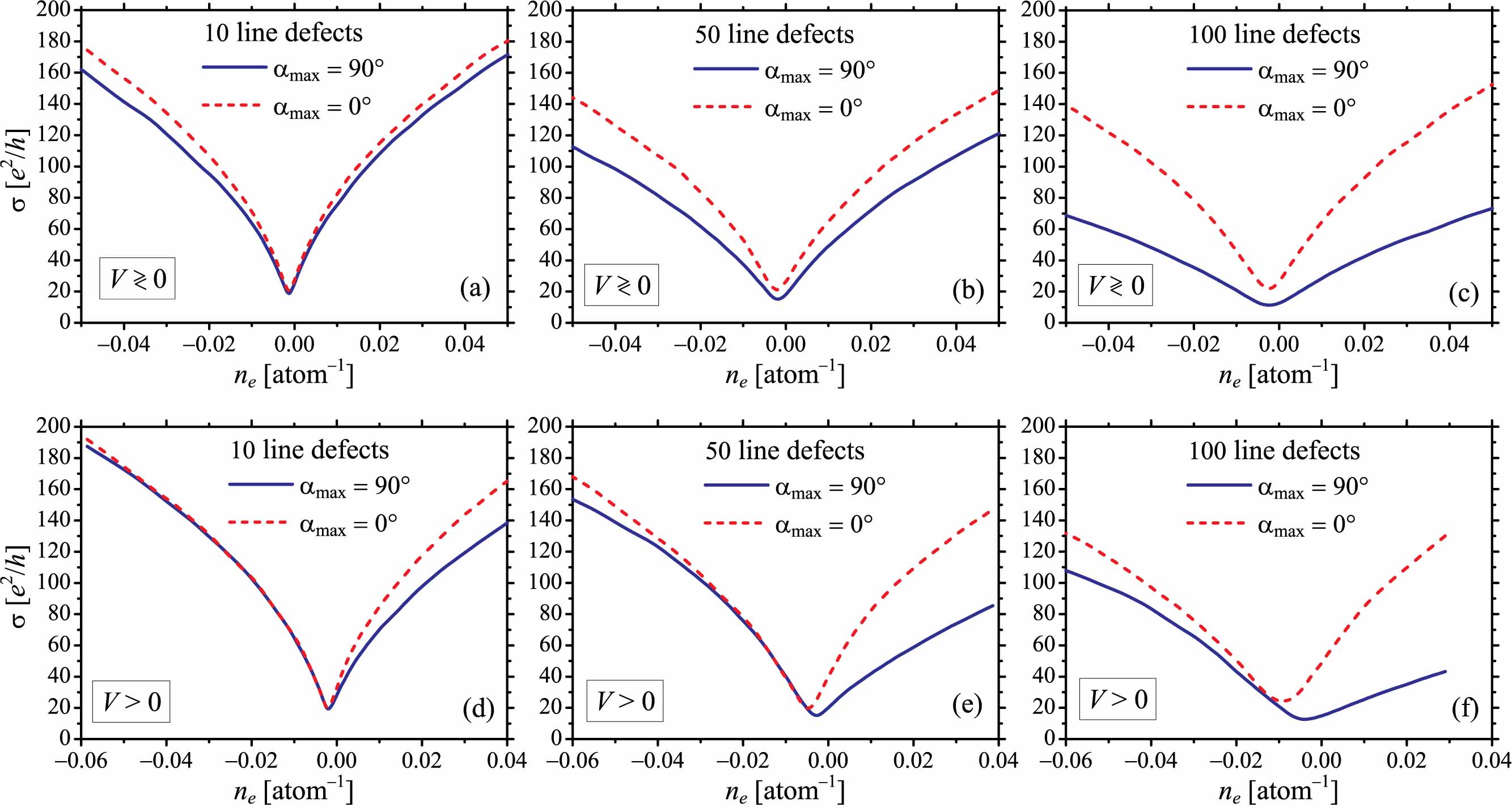}

\caption{(Color online) Conductivities $\sigma_{xx}^{\alpha\mathrm{_{max}}}$
vs. the relative charge carrier (electron) density for different number
(10, 50, 100) of random ($\alpha_{\textrm{max}}=90^{\circ}$) and
parallel ($\alpha_{\textrm{max}}=0^{\circ}$) in each realization
lines for (a)--(c) symmetric ($V\gtrless0$) and (d)--(f) asymmetric
($V>0$) scattering potentials (with $\triangle=0.25u$). Each curve
is averaged over 20 different configurations of lines (including their
orientations and distances between them).}

\label{Fig_Cond_10-50-100_lines} 
\end{figure*}

In Figure \ref{Fig_Cond_10-50-100_lines}, we show the electron-density
dependence of the conductivity of graphene sheets with different (10,
50, 100) number of linear defects for the cases of symmetric and asymmetric
potentials. First, for a given defect concentration the conductivity
of graphene with the correlated line defects, $\alpha\mathrm{_{max}}=0^{\circ}$,
increases in comparison to the case of uncorrelated defects, $\alpha\mathrm{_{max}}=90^{\circ}$
(see Fig. \ref{Fig_Cond_10-50-100_lines}). This can be contrasted
with the case of point defects,when the correlation in the defect
position practically does not affect the conductivity (Fig. \ref{Fig_Cond_corr}).
Second, for a given electron density, the relative increase of the
conductivity for the case of fully correlated line defects in comparison
to the case of uncorrelated ones is higher for a larger defect density.
This is an expected result since correlation effect manifests itself
stronger for a larger number of objects-to-be-correlated---line defects
at hand.

Finally, note some features that show the obtained dependencies $\sigma=\sigma(n_{e})$
in (CVD and epitaxial) graphene, where the charged line-acting defects
are believed to represent the limiting scattering mechanism.

First, the conductivity exhibits a pronounced sublinear electron-density
dependence and depends weakly on the Thomas--Fermi screening wavelength
\cite{Radchenko et al. 2}. Our numerical calculations are consistent
with the recent experimental results \cite{Kim2009,Song2012,Tsen2010}
that also exhibit sublinear density dependence of $\sigma$. This
provides an evidence in support that the line defects represent the
dominant scattering mechanism in both CVD and epitaxial graphene \cite{Kim2009,FerreiraEPL,Song2012}.

Second, the conductivities of samples with different line configurations
exhibit significant variations between each other \cite{Radchenko et al. 2}.
This is in strong contrast to the case of short- and long-range point
scatterers where corresponding conductivities of samples of the same
size and impurity concentrations practically did not show any noticeable
differences for different impurity configurations \cite{Radchenko et al. 1}.
We attribute this to the fact that in contrast to point defects, the
line defects are characterized not only by their positions, but also
by directions (orientations) and their intersections as well. Such
additional characteristics result in much more possible distributions
of the potential which, in turn, leads to the differences in the conductivity
curves.

Third, for the symmetric potential the conductivity curves are symmetric
with respect to the neutrality (Dirac) point, while the asymmetric
one leads to the asymmetry in the conductivity, cf. Figs.\ref{Fig_Cond_10-50-100_lines}(a)--(c)
and (d)--(f). Such the asymmetry between the holes and electrons,
being also seen in transport calculations in graphene with point \cite{Radchenko et al. 1,Leconte11,Robinson2008,Wehling2010}
and line \cite{Radchenko et al. 2,FerreiraEPL} defects, causes a
quantitatively different conductivity enhancements for the orientationally-correlated
line defects as Fig. \ref{Fig_Cond_10-50-100_lines} demonstrates.

\begin{figure*}[!t]
\includegraphics[width=1\textwidth]{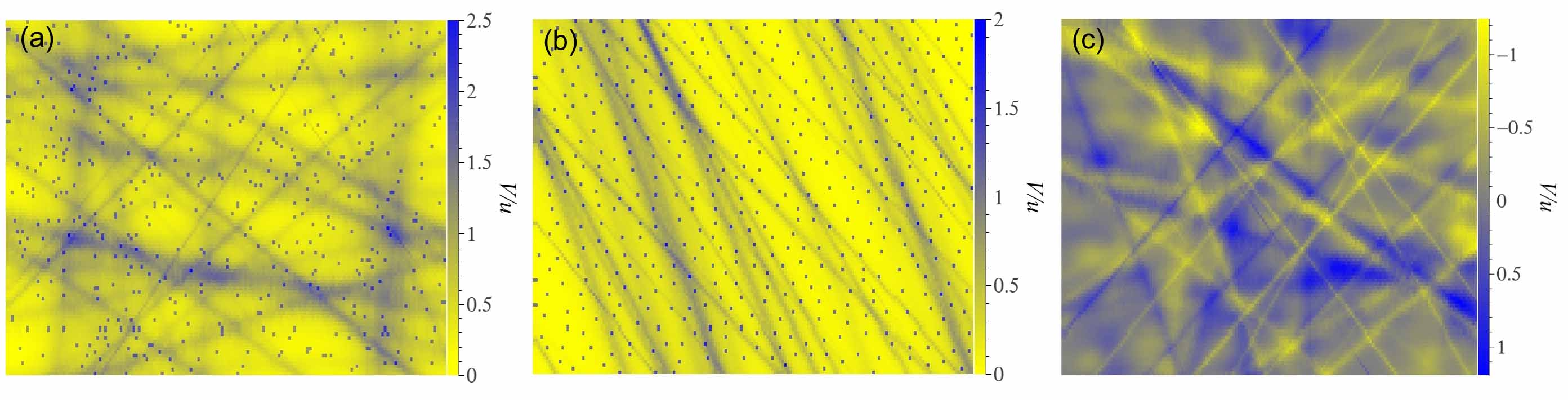}

\caption{(Color online) Distribution of scattering potential in graphene with
a simultaneous presence of random (a), (c) and correlated (b) point
and line defects, where the point ones (impurities) are short-range
weak (a), (b) and long-range Gaussian (c).}

\label{Fig_Potential_lines_points} 
\end{figure*}

\begin{figure*}[!t]
\includegraphics[width=1\textwidth]{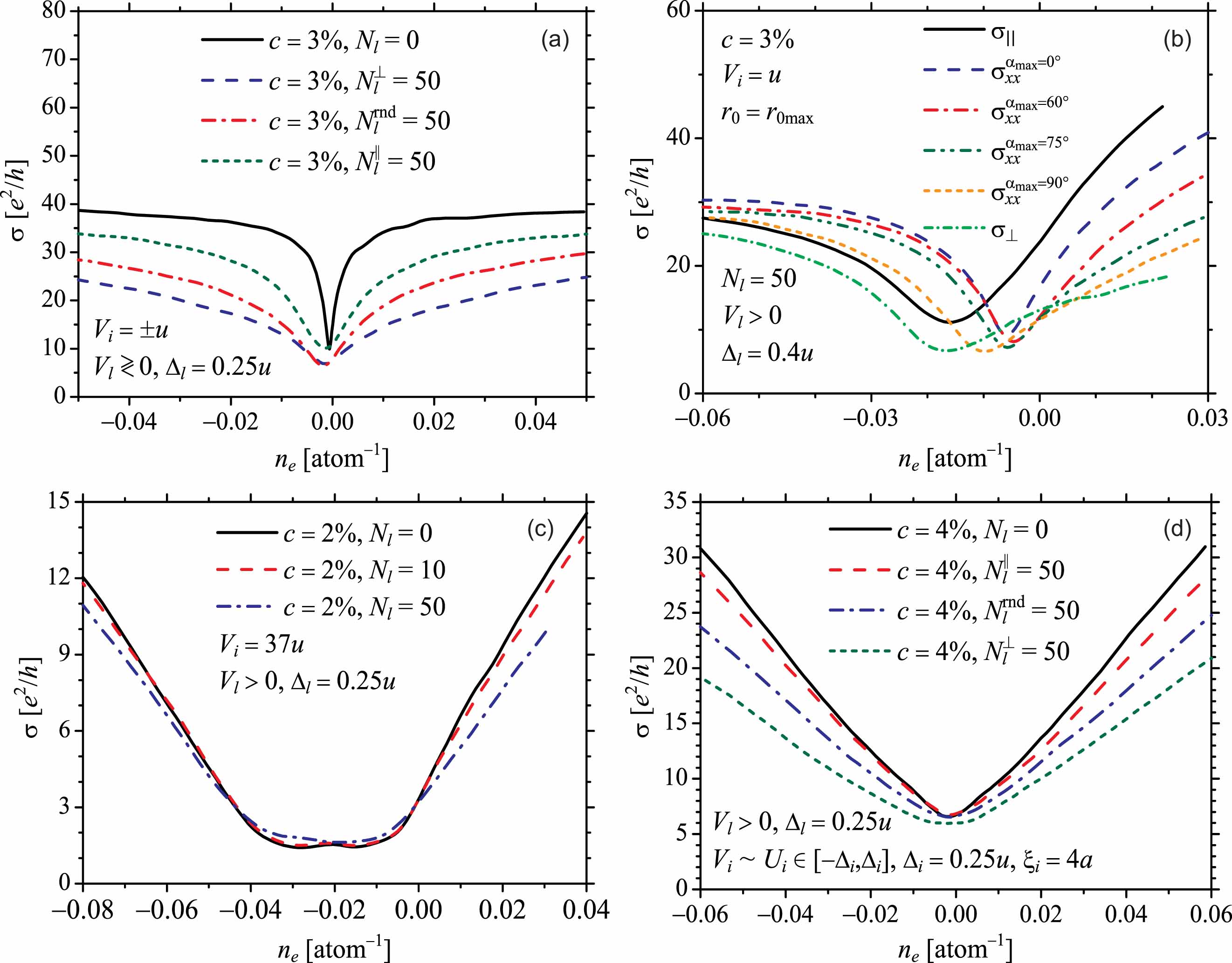}

\caption{(Color online) Conductivity vs. the electron concentration in graphene
with both point and line defects. Subscripts $i$ and $l$ denote
impurities and lines, respectively; $N$ is a number of random, parallel
($\parallel$) or perpendicular ($\perp$) lines. Impurities are short-range
weak symmetric (a) or asymmetric (b), short-range strong (c), and
long-range Gaussian (d). }

\label{Fig_Point_lines} 
\end{figure*}

In conclusion of the section note that the presence of both defect
types, point and line ones (Fig. \ref{Fig_Potential_lines_points}),
which seems even more realistic than all cases considered above, can
significantly affects the behavior of conductivity in comparison with
the case when only one type of them is considered (Fig. \ref{Fig_Point_lines}).
Conductivity for short-range weak impurities, being electron-density
independent, becomes sublinear at the addition of charged line defects
to them {[}Fig. \ref{Fig_Point_lines}(a){]}. An interplay between
the point and line defects, being modeled by the potential of the
same {[}e.g., positive as in Fig. \ref{Fig_Point_lines}(b){]} sign,
can suppress the electron\textendash{}hole asymmetry revealed if they
are taken into account separately (see Figs. \ref{Fig_Cond_corr},
\ref{Fig_DOS_Cond_ord}, and \ref{Fig_Cond_10-50-100_lines}). However,
an addition of the line defects to the short-range strong impurities
weakly change the $\sigma=\sigma(n_{e})$ dependence {[}Fig. \ref{Fig_Point_lines}(c){]}
due to essentially-different scattering forces of the potentials ($V_{i}\gg V_{l}$).
At last, an addition of the charged line-acting defects to the long-range
Gaussian ones, remains the density dependence to be linear {[}Fig.
\ref{Fig_Point_lines}(d){]} in spite of its robust sublinear dependence
for line defects without point ones.

\section*{Effect of Nitrogen or Boron Doping Configurations in Graphene: DFT
vs. Kubo--Greenwood Formalism}

\begin{figure*}[!t]
\includegraphics[width=1\textwidth]{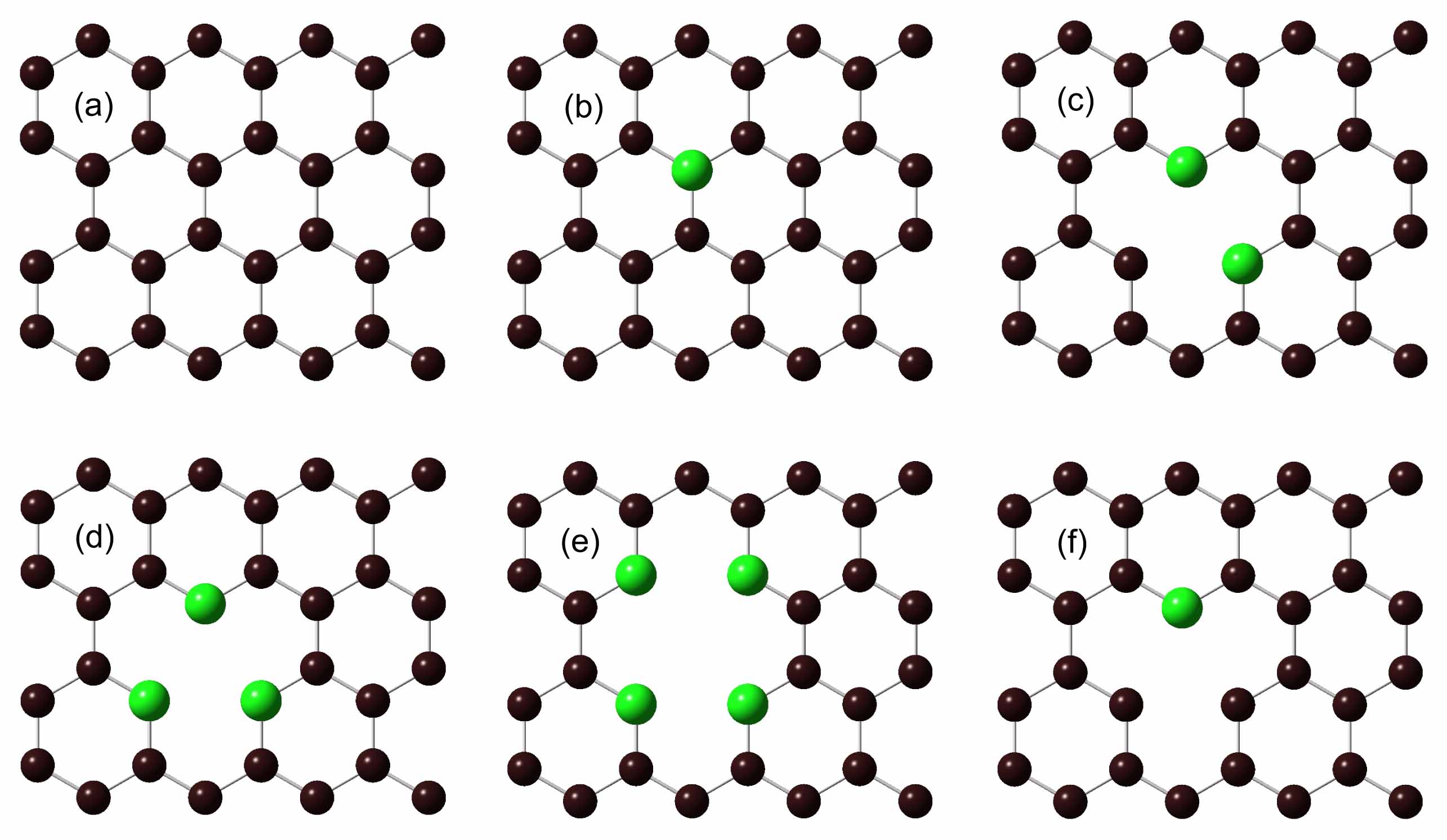}

\caption{(Color online) Atomic configurations of the pure (a) and N- or B-doped
graphene samples with substitutional defect (b), dimerized (c), trimerized
(d), tetramerized (e), and monomeric (f) pyridine-type defects. }

\label{Fig_B(N)_configurations} 
\end{figure*}

\begin{figure*}[!t]
\includegraphics[width=1\textwidth]{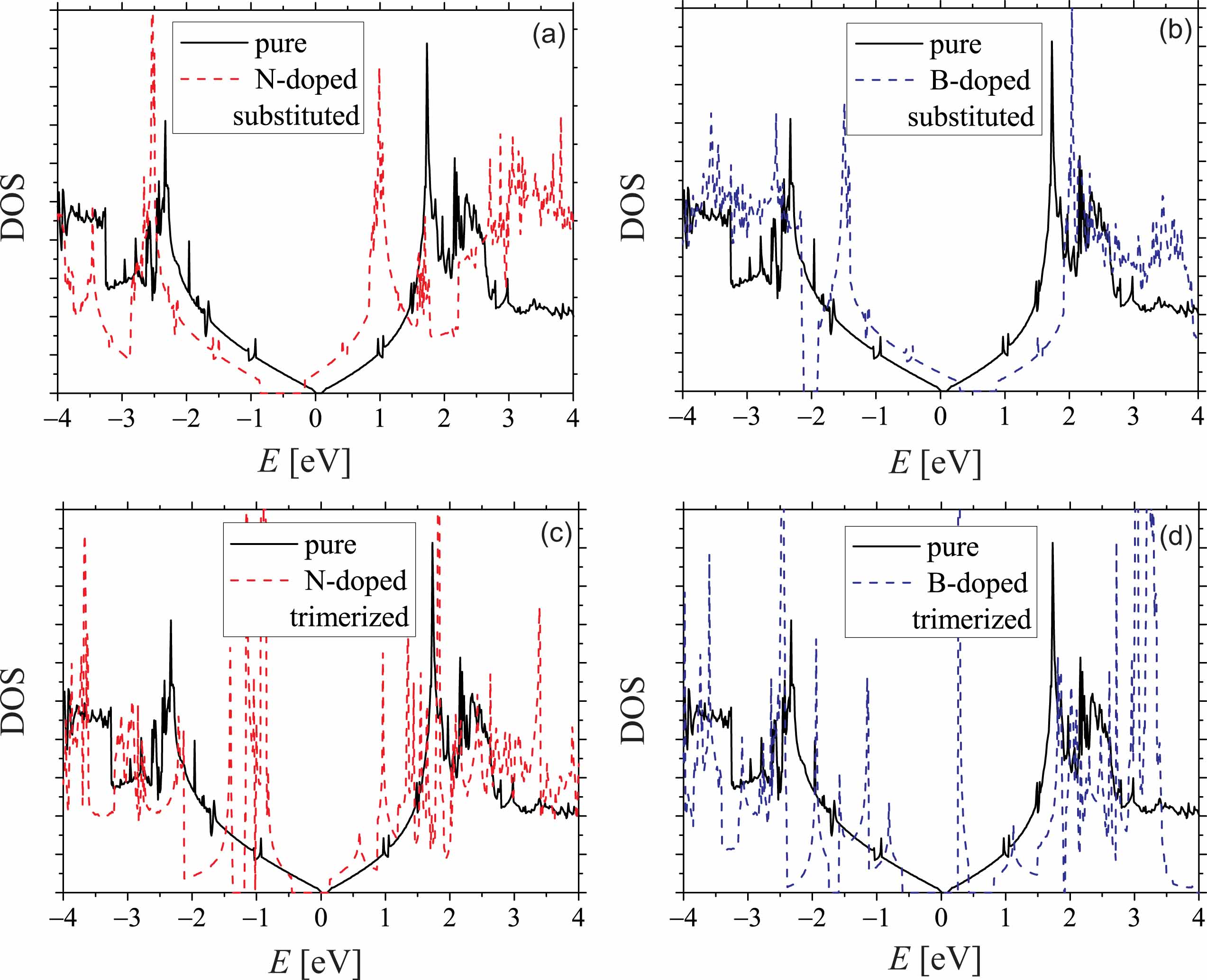}

\caption{(Color online) Density of states for graphite- (a), (b) and pyridine-like
(b), (d) substitutions in N- (a), (c) and B-doped (b), (d) graphene. }

\label{Fig_DOS-DFT} 
\end{figure*}

\begin{figure*}[!t]
\includegraphics[width=1\textwidth]{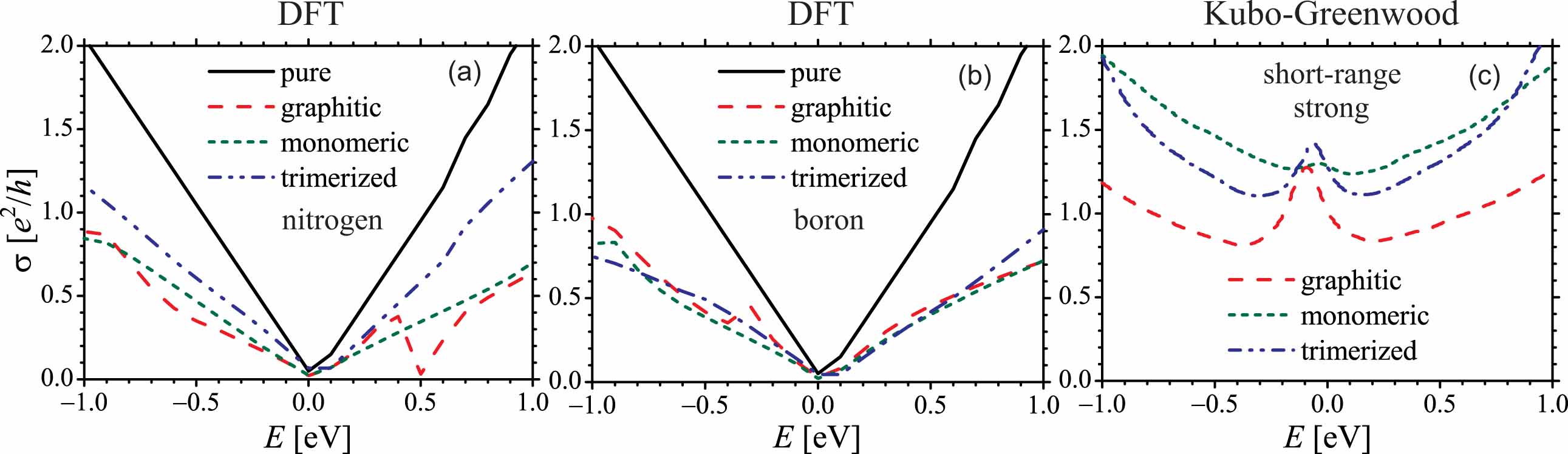}

\caption{(Color online) Conductivity vs. the Fermi energy calculated within
the DFT (a), (b) and Kubo--Greenwood (c) methods for graphene with
12.5\% of nitrogen (a), boron (b), and short-range strong impurities
(c). }

\label{Fig_Cond_DFT-Kubo} 
\end{figure*}

The present section deals with a comparative implementation of the
above-described Kubo--Greenwood formalism and the density-functional-theory-based
(DFT) method to calculate the charge transport in a B- or N-doped
graphene samples. Boron and nitrogen are the most suitable and therefore
commonly used substitutional dopants for incorporation into the graphene
lattice. Recent experimental measurements of x-ray photoelectron spectra
of N-doped nanotubes and graphene have revealed the presence of several
N-doped configurations \cite{Wei2009,Lin2010,Ayala2007,Guo2010}.
Neglecting here the cases of topology changes in the relative positions
of the honeycomb-lattice sites (i.e. line defects), there are five
configurations for N- or B-doped graphene computational domain. Their
geometries are illustrated in Fig. \ref{Fig_B(N)_configurations}.
There are one graphite-type defect, N or B substitution in Fig. \ref{Fig_B(N)_configurations}(b),
and four pyridine types of defects: dimerized in Fig. \ref{Fig_B(N)_configurations}(c),
trimerized in Fig. \ref{Fig_B(N)_configurations}(d), tetramerized
in Fig. \ref{Fig_B(N)_configurations}(e), and monomeric in Fig. \ref{Fig_B(N)_configurations}(f).
We classify these configurations into two groups: point defects {[}single
dopant atoms (or vacancies), as shown in Fig. \ref{Fig_B(N)_configurations}(b){]},
and complex ones containing both substitutional impurity atoms and
vacancies arranged in a fixed clusters distributed over all structure
{[}Figs. \ref{Fig_B(N)_configurations}(c)--(f){]}.

Being traditionally a powerful tool for the study of electronic and
transport properties of materials, in contrast to the Kubo--Greenwood
approach, the DFT method, however, does not allow to treat a large
graphene systems. Here, we chose the origin graphene structure (cluster)
consisting of 32 sites {[}Fig. \ref{Fig_B(N)_configurations}(a){]}
composing a rectangular supercell of the $8.5\times9.8$ Å size. To
exclude the interaction with other graphene sheets, the given supercell
is wrapped in a vacuum of 17 Å of thickness along the $y$-axe. The
QUANTUM ESPRESSO computational packet \cite{Giannozzi} was used for
calculations within the electron density functional method. To describe
the exchange-correlation energy, we used the LDA approach in the Perdew--Zunger
parametrization (BLYP for density of states calculation). C, N, and
B atoms are described by the corresponding pseudopotentials US-PP
\cite{pseudopotentials}. Separation kinetic energy for the wave functions
and charge densities are 30 and 300 Ry, respectively. Transport calculations
are carried out using the PWCOND codes \cite{Smogunov}.

As evidenced by the DFT-based calculations of the DOS (Fig. \ref{Fig_DOS-DFT}),
the effect of boron and nitrogen impurities is symmetrical with respect
to the Dirac point. Incorporation of boron or nitrogen atoms in substitution
within the carbon matrix gives rise to the efficient $p$- or $n$-type
doping of graphene. Oscillations in DOS for pyridine-type defects
{[}Figs. \ref{Fig_DOS-DFT}(c) and (d){]} become stronger in comparison
with DOS for the graphite-type defects {[}Figs. \ref{Fig_DOS-DFT}(a)
and (b){]}, that may be associated with a vacancy effect. Note that
the oscillations, present in the DOS computed from DFT (Fig. \ref{Fig_DOS-DFT}),
are smoothed in the DOS from the Kubo--Greenwood approach (see DOS
for random impurities in Fig. \ref{Fig_DOS_Cond_ord}) due to the
significantly larger computational domain treated within the framework
of the Kubo method.

Conductivity in Fig. \ref{Fig_Cond_DFT-Kubo} is calculated within
the both DFT and Kubo--Greenwood approach, where in the latter, impurities
are modeled by the short-range strong scattering potential. The DFT
calculations result mainly to the linear energy dependence of the
conductivity {[}Figs. \ref{Fig_Cond_DFT-Kubo}(a) and (b){]}, it means
that electron-density dependence $\sigma=\sigma(n_{e})$ should be
sublinear since $E\propto\sqrt{n_{e}}$ \cite{Radchenko et al. 1,Radchenko et al. 2}.
However, from the Kubo calculations, $\sigma=\sigma(n_{e})$ is obtained
to be close to linear {[}see Figs. \ref{Fig_Cond_corr}(c) and \ref{Fig_Cond_DFT-Kubo}(c){]}.
Another difference is that $\sigma_{\mathrm{min}}$ obtained by DFT
method is much more smaller than that obtained from the Kubo method.
Also, results in Fig. \ref{Fig_Cond_DFT-Kubo}(c) are evidence of
the fact that the short-range strong impurities manifest themselves
as the stronger scatterers as compared with vacancy ones, but this
is not the case obtained from DFT {[}Figs. \ref{Fig_Cond_DFT-Kubo}(a),
(b){]}.

\section*{Conclusion}

The statistical-thermodynamics and kinetics models of both substitutional
and interstitial atomic order in the two-dimensional graphene-based
crystal lattices for a wide interval of stoichiometries are constructed.
Ordered distributions of substitutional and interstitial atoms over
the sites and interstices of the honeycomb lattice at the different
compositions and temperatures are predicted and described theoretically.
The ranges of values of interatomic-interaction parameters providing
the low-temperature superstructural stability are determined within
the framework of both the third-nearest-neighbor Ising model and,
more realistic, model taking into account interactions of all atoms
present in the system at hand. The first model results in the instability
of some predicted superstructures, while the second one shows that
all predicted superstructures are stable at the certain values of
interatomic-interaction energies. Even short-range interatomic interactions
provide a stability of some graphene-based superstructures, while
only long-range interactions stabilize others. Inasmuch as the intrasublattice
and intersublattice interchange (mixing) energies are competitively
different with the dominance of the latter, the long-range atomic
order parameter(s) may relax to the equilibrium value(s) nonmonotonically.

A numerical study of electronic transport in single-layer graphene
is performed by means of an efficient time-dependent real-space Kubo--Greenwood
approach, which is especially suited to treat large graphene systems
containing millions of atoms. The presence of neutral and/or charged
point and/or line defects in graphene is modeled by various short-
and long-range scattering potentials. The strong short-range scattering
potential describes neutral adatoms covalently bond to graphene. The
long-range Gaussian-shaped potential is appropriated for screened
charged impurities on graphene and/or dielectric substrate surface.
The self-consistent Thomas--Fermi approximation-based effective potential
is used for charged line-acting defects (grain boundaries in CVD-grown
polycrystalline graphene, atomic substrates in epitaxial graphene,
etc).

Correlation in the distribution of impurity atoms gives a slight rise
(up to $30\%$) in the conductivity only for the case of weak short-range
potential and only if it is asymmetric (repulsive). In other the most
experimentally relevant cases, namely, the short-range strong and
long-range Gaussian scatterers, correlation does not affect the conductivity.

Ordering of impurities can give rise to conductivity up to tens times
for weak and strong short-range scatterers as compared with the case
when dopants are distributed randomly. However, as for the correlation,
ordering does not affect the conductivity for the long-range-acting
Gaussian potential.

Studying numerically the charge carrier transport in graphene with
one-dimensional charged defects, we got electron-density dependencies
of the conductivity, which showed some new features as compared with
those obtained in case of point defects. First, the conductivity is
found to be a robust sublinear function of electronic density and
weakly dependent on the Thomas--Fermi screening wavelength. The calculated
sublinear density dependence for the case of linear defects is quite
different from the case of short- and long-range point scatterers,
where the numerical calculations show a density dependence close to
linear. We attribute the atypical, but consistent with recent experimental
reports, behavior of conductivity to the extended nature \textit{\emph{of
one-dimensional}} charged defects. Second, the conductivities of samples
with different impurity geometries exhibit significant variations
between each other. This is due to the fact that in contrast to point
defects, the line defects are characterized not only by their positions,
but also by directions (orientations) and their intersections as well.
Such additional characteristics result in much more possible distributions
of the potential which, in turn, leads to the differences in the conductivity
curves.

The anisotropy of the conductivity along and across the line defects
is revealed, which agrees with the experimental measurements for CVD
graphene grown on Cu and epitaxial graphene grown on SiC. For a given
concentration of the line defects, the conductivity of graphene with
orientationally-correlated defects increases in comparison to the
case of the uncorrelated line defects. For a given electron density,
a relative increase of the conductivity for the case of fully correlated
line defects in comparison to the case of uncorrelated defects is
higher for a larger defect density.

A simultaneous account of both point and line defects can qualitatively
and quantitatively affect the conductivity behavior in comparison
with the case when only one type of them is considered. An interplay
between the point and line scatterers modeled by the potential of
the same sign suppresses the electron--hole asymmetry revealed if
they are taken into account separately. If both point and line defects
are correlated and/or ordered, it can give rise in the conductivity
of graphene up to hundreds times vs. their random distribution, and
thereby can serve as an additional tool to control and govern the
transport properties in graphene.

\subsection*{Acknowledgments}

T.M.R. benefited immensely from collaboration with Igor Zozoulenko,
Artsem Shylau, Aires Ferreira, and appreciates discussions with Stephan
Roche and Sergei Sharapov. 

\label{lastpage-01} 
\end{document}